\documentclass[a4paper,12pt,preprint]{article}

\usepackage{etex}
\usepackage[intlimits]{amsmath}
\usepackage{hyperref}
\usepackage[russian,english]{babel}
\usepackage{euscript}
\usepackage{amssymb,amsmath,amsthm}

\usepackage[all]{xy}
\usepackage{bbm}
\usepackage{esvect}
\usepackage{setspace}
\usepackage{epsfig}
\usepackage{caption}
\usepackage{wrapfig}
\usepackage[textwidth=3cm,backgroundcolor=blue!10]{todonotes}
\usepackage{youngtab}
\usepackage{empheq}

\usepackage{array}

\usepackage{pxfonts}
\usepackage{txfonts}
\usepackage[T1]{fontenc}

\usepackage{pstricks,pst-node}
\usepackage{color}
\usepackage{pifont,bbding,ifsym}
\usepackage{wasysym}

\DeclareFontFamily{U}{mathx}{\hyphenchar\font45}
\DeclareFontShape{U}{mathx}{m}{n}{
      <5> <6> <7> <8> <9> <10>
      <10.95> <12> <14.4> <17.28> <20.74> <24.88>
      mathx10
      }{}
\DeclareSymbolFont{mathx}{U}{mathx}{m}{n}
\DeclareFontSubstitution{U}{mathx}{m}{n}
\DeclareMathAccent{\widebar}{0}{mathx}{"73}

\newcommand{\bea}{\begin{equation}}
\newcommand{\eea}{\end{equation}}
\newcommand{\bear}{\begin{eqnarray}}
\newcommand{\eear}{\end{eqnarray}}
\newcommand{\bearr}{\begin{eqnarray*}}
\newcommand{\eearr}{\end{eqnarray*}}

\newcommand{\beal}{\begin{align}}
\newcommand{\eeal}{\end{align}}
\newcommand{\beall}{\begin{align*}}
\newcommand{\eeall}{\end{align*}}

\newcommand{\tr}{\mathrm{tr}\,}
\newcommand{\CP}{\mathbf{C}\mathrm{P}}
\newcommand{\CC}{\mathbf{C}}
\newcommand{\dd}{\partial}
\newcommand{\im}{\mathrm{i}\,}

\DeclareMathAlphabet{\mathpzc}{OT1}{pzc}{m}{it}
\newcommand{\comment}[1]{}

\newcommand{\MM}{M}
\newcommand{\LM}{\mathcal{L}M}
\newcommand{\fl}{\mathcal{F}}
\newcommand{\Zz}{\mathbf{Z}}
\newcommand{\Hh}{\mathrm{H}}
\definecolor{lightgrey}{RGB}{200,200,200}
\newcommand{\0}{\textcolor{lightgrey}{0}}
\newcommand{\bl}{\boldsymbol\lambda}

\makeatletter
\def\@seccntformat#1{\@ifundefined{#1@cntformat}%
{\csname the#1\endcsname\quad}
{\csname #1@cntformat\endcsname}
}
\def\section@cntformat{{\normalfont\large\thesection.}\quad}
\def\subsection@cntformat{\textsection\, \thesubsection.\quad}
\def\subsubsection@cntformat{\textsection\textsection\, \thesubsubsection.\quad}
\makeatother

\newcommand{\ssection}[1]{%
  \section[#1]{\centering\normalfont\scshape #1}}
\newcommand{\ssubsection}[1]{%
   \subsection[#1]{\raggedright\normalfont  #1}}
\newcommand{\ssubsubsection}[1]{%
   \subsubsection[#1]{\raggedright\normalfont  #1}}

\usepackage{textpos}
\begin{document}

\title {The geometry of antiferromagnetic spin chains}
\author {Dmitri Bykov\footnote{Emails:
dbykov@nordita.org, dbykov@mi.ras.ru}
\\
{\small{\it Nordita, Roslagstullsbacken 23, 106 91 Stockholm, Sweden}} \\ {\small {\it Steklov
Mathematical Institute, Gubkina str. 8, 119991 Moscow, Russia \;}}}
\date{}
\maketitle 
\vspace{-0.75cm}
\begin{abstract}
We construct spin chains that describe relativistic $\sigma$-models in the continuum limit, using symplectic geometry as a main tool. The target space can be an arbitrary complex flag manifold, and we find universal expressions for the metric and $\theta$-term.
\end{abstract}

\begin{textblock}{4}(11.3,-9)
\underline{\small NORDITA-2012-47}
\end{textblock}

\begin{flushright}
\emph{The true goal is not to reach the uttermost limits, \\but to discover a completeness that knows no boundaries.}

Rabindranath Tagore
\end{flushright}

\ssection{Introduction}

The application of coherent states in the physics of spin chains is a beautiful subject, whose physics and mathematics sides are both extremely rich. The aim of this paper is to apply the corresponding mathematical formalism to a description of long-range excitations around antiferromagnetic vacua of particular spin chains with $U(N)$ symmetry in the quasiclassical (large spin) limit. The peculiarity of these spin chains is that the resulting continuum model is nothing but a $\sigma$-model with target space a flag manifold. More concretely, we propose that the $\sigma$-model with target space $\frac{U(N)}{U(n_1)\times\cdots \times U(n_m)}$ ($\sum\limits_{i=1}^m n_i=N$) can be obtained from a spin chain with the following Hamiltonian:
\bea\label{1}
\mathcal{H}=\sum\limits_{i=1}^{L} \, \sum\limits_{k=1}^{m-1} \, d_k\; \vec{S}_i\cdot \vec{S}_{i+k},
\quad\quad
d_k=\sqrt{\frac{m-k}{k}}
\eea
In the above formula $\vec{S}_i$ represent the generators of a representation of $\mathfrak{su}_N$, sitting at site $i$. The representations at $m$ consecutive sites can be described via Young diagrams that consist of single columns of height $n_1, \cdots, n_m$, and a permutation of these sites generically produces a $\sigma$-model with the same target space, but with a different metric and $\theta$-term.

The actual construction that produces the above result relies on methods from representation theory and, even more importantly, symplectic geometry. Therefore we postpone the derivation of the results to Sections \ref{metricsec}, \ref{topsec} and begin this paper by giving an overview of the mathematical formalism. One faces the necessity for this formalism as soon as one embarks on the construction of a path integral representation for a spin chain, which, in turn, is the most natural framework for the quasiclassical and continuum limits.

The construction of the spin chain path integral can be roughly separated in two stages: the kinematical and dynamical parts. The kinematical part, which is the subject of Section \ref{pathint}, has to do with the description of the phase space of a single spin --- this is essentially a part of representation theory, its aim being the description of coherent states in a given representation of the global symmetry group $G$. From the mathematical viewpoint, this is an etude in the so-called Borel-Weil-Bott theorem, reviewed in Section \ref{BWB} of the present paper. Suppose $G$ is a group of  linear automorphisms of a vector space $V$. Quite generally, the manifold of coherent states for any representation is a certain manifold of linear flags in $V$, which can be viewed alternatively as an orbit of $G$ on the space of its coadjoint representation. This statement allows one to make a connection to the method of orbits in representation theory, and this is indeed necessary for the construction of the path integral. It turns out that the kinetic part of the classical action for the single spin is in fact given by a particular symplectic form on the respective coadjoint orbit. Therefore the solution of the kinematical part of the problem may be viewed as a fruit of the Borel-Weil-Bott theorem and the orbit method.

\hspace{1cm}\begin{pspicture}(15,5)
\psset{linecolor=blue} 

\psbezier[linewidth=1.5pt,linestyle=dashed,linecolor=orange]{->}(9,3.9)(8,3.5)(7,3.2)(6.2,3.1)
\psbezier[linewidth=1.5pt,linestyle=dashed,linecolor=orange]{->}(10,3)(9,2.6)(6,2.5)(5.7,2.5)
\psbezier[linewidth=1.5pt,linestyle=dashed,linecolor=orange]{->}(8.1,1.3)(7,1.7)(6,1.8)(5.3,1.9)

\psbezier[linewidth=1.5pt,linestyle=dashed,linecolor=orange]{->}(2.6,1.3)(3,1.7)(4,1.8)(5.3,1.9)
\psbezier[linewidth=1.5pt,linestyle=dashed,linecolor=orange]{->}(2.5,3)(3.4,2.7)(4.5,2.65)(5.7,2.5)
\psbezier[linewidth=1.5pt,linestyle=dashed,linecolor=orange]{->}(5.2,3.9)(5,3.6)(5.7,3.2)(6.2,3.1)

\psline[linewidth=1.5pt,linecolor=orange]{->>}(8,2.635)(7.7,2.6)

\psbezier[linewidth=2pt,linecolor=red](5,0.75)(5,2.5)(7,3.5)(7,4.1)
\psccurve[linewidth=2pt,fillstyle=none](2,2)(10,3)(6,4)
\psline[linewidth=1.5pt]{->}(5.57,2.32)(4.4,3.1)
\psline[linewidth=1.5pt]{->}(5.57,2.32)(6.25,3.4)

\rput(4.3,2.9){$p$}
\rput(5.95,3.4){$q$}
\rput(4.8,1.5){$L$}
\rput(8.5,2){$M$}
\rput[color=orange](8,2.9){$I$}
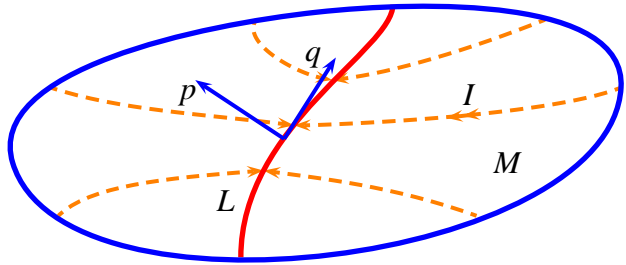
\rput(6,0.25){Fig. 1. The topography of $M$ in the vicinity of $L$.}
\end{pspicture}

The dynamical part, described in Section \ref{dynamical}, in turn describes the interactions of spins sitting at various sites. Speaking prosaically, it is all about the choice of a Hamiltonian for the spin chain. However, certainly some Hamiltonians are ``better'' than others in the sense that they lead to beautiful geometrical structures. We illustrate this with the example of a continuum limit of a certain spin chain. The peculiarity of this spin chain, which is important for this construction to work, is that the minimum manifold $L$ of the Hamiltonian can be viewed as the locus of zeros of a certain moment map, $L\simeq \mu^{-1}(0)$, and, moreover, it is a single $G$-orbit. In Section \ref{orbtheo} we explain, in what sense this situation is special, providing the necessary background material from symplectic geometry. Once these requirements are fulfilled we show that the continuum limit of this spin chain results in a two-dimensional \emph{relativistic} sigma-model with target space $L$. Moreover, the resulting Lagrangian of the sigma model can be described in a general setup. It turns out that the metric on $L$ can be obtained by a rather universal geometric construction. The topological $\theta$-term is in turn severely restricted by the translational invariance of the spin chain and can be described in a simple way in terms of certain canonical generators. One of the interesting consequences of this result is that a simple permutation of sites of the spin chain generically leads to a different $\theta$-term. In a sense, this is a way to physically realize the generators of the cohomology group $\Hh^2(L, \mathbf{Z}_m)$\footnote{$m$ refers to the number of factors in the denominator of the coset $L={U(N)\over U(n_1)\times \cdots \times U(n_m)}$}! According to the argument of Haldane \cite{H}, this term is related to the absence or presence of a mass gap in the spin chain, and is therefore of crucial importance.

The appendices offer derivations of the results presented in Sections \ref{metricsec} and \ref{topsec} of the paper, as well as an example of integration over a flag manifold and an example of calculation of a quadratic Casimir using oscillator algebra.

This paper is in some sense a continuation of \cite{Bykov}. Some familiarity with that paper will certainly be useful, in particular in order to understand the logic of our manipulations with the spin chain Lagrangian in Sections \ref{hamsec}, \ref{metricsec}, \ref{topsec} (although all calculations are given in Appendix \ref{A}). A large source of inspiration for this work is the paper of F.A.Berezin \cite{Berezin}, who was probably one of the first to introduce geometry into quantization in the sense used in this paper, and the much more recent paper of E.Witten \cite{witten}. Important work on the mathematical description of coherent states was done by A.M.Perelomov, see \cite{Perelomov}. Substantial work on the subject of Haldane continuum limits was done by I.Affleck, see \cite{Affleck} as an example.

\ssection{Path integrals for spin chains}\label{pathint}


The goal of this Section is to review a general construction of path integral representations for spin chain partition functions. Roughly speaking, we are aiming at obtaining an expression of the following sort:
\begin{empheq}[box=\fbox]{align}
\label{evopxxx}
\hspace{1em}
\mathcal{Z} \equiv \tr(e^{-\beta \mathcal{H}_X})= \int\; \prod\limits_{i, \,t\in [0,1]} d\mu(z_i(t),\bar{z}_i(t))\;\exp{(-\mathcal{S})}, \; \textrm{where} \hspace{2em}\\
\label{actionxxx}
\;\;\;z_i\in \CP^{N-1}\quad\textrm{and}\quad\mathcal{S} = m\;\int\limits_0^1\,dt\, \sum\limits_i \left( i \frac{\dot{z}_i \circ \bar{z}_i}{z_i\circ \bar{z}_i} +
\beta \frac{z_{i}\circ  \bar{z}_{i+1}}{z_{i}\circ \bar{z}_{i}} \frac{z_{i+1}\circ \bar{z}_{i}}{z_{i+1}\circ \bar{z}_{i+1}} \right)
\hspace{1em}
\end{empheq}
This formula is for a nearest neighbor spin-spin coupling described by the Hamiltonian $\mathcal{H}_{X}=\sum\limits_i\,\vec{S}_i\cdot \vec{S}_{i+1}$ with $SU(N)$ symmetry; $m$ is a positive integer indicating the representation at each site, i.e. it is the $m$-th symmetric power of the fundamental. In what follows we will discuss various generalizations, both of the kinetic term (Sections \ref{BWB}, \ref{cohstates}, \ref{circaction}) and of the Hamiltonian (Section \ref{dynamical}).

\ssubsection{The kinematical aspect. The Borel-Weil-Bott theorem.}\label{BWB}

Suppose $H$ is a spin chain Hamiltonian with symmetry group $U(N)$. In this Section we explain how one can write an expression for it in terms of the so-called coherent states. In order to accomplish this task one first needs to find out what the coherent states are for a given site of the spin chain. There is a very general theorem that gives an answer to this question, which is usually attributed to Borel, Weil and Bott (BWB). It gives in fact a complete geometric (and therefore beautiful) description of the whole representation theory of $U(N)$ (it is also generalizable to other Lie groups, but we prefer to focus here on this simplest example). It goes as follows.

The assertion of the BWB theorem is that a finite-dimensional representation of $U(N)$ with highest weight $\vec{\lambda}$ can be modeled on the space of holomorphic sections of a holomorphic line bundle over a complete flag manifold
\bea\label{fullflag}
\fl_N=U(N)/U(1)^N .
\eea
The line bundle is commonly denoted $L_{\vec{\lambda}}$. Morally speaking, one can think of these sections as (not uniquely defined) functions $f_i(z)$ on $\fl_N$, which transform according to the representation $\tau$ under the action of $G=U(N)$:
\bea
f_i(g\circ z)=\sum\limits_{j=1}^{\mathrm{dim}\,\tau}\;\tau(g)_i^j\;f_j(z)
\eea
So how is the line bundle $L_{\vec{\lambda}}$ built? In order to understand this, first of all one has to know the second cohomology of the flag manifold:
\bea
\mathrm{H}^2(\fl_N, \Zz)=\Zz^{N-1},
\eea
therefore there are $N-1$ linearly independent 2-forms, that are the generators of $\Hh^2(\fl_N)$. As a model for $\Hh^2(\fl)$ we will use the following. On $\fl_N$ there are $N$ standard (or tautological) line bundles, $L_1, \cdots , L_N$, their sum being trivial:
\bea\label{triv}
\overset{N}{\underset{i=1}{\oplus}}\;L_i = \fl_N\, \times \CC^N\; .
\eea
Their first Chern classes provide us with $N$ closed 2-forms: $\Omega_i = c_1(L_i), \; i=1\cdots N$.  Due to the property (\ref{triv}) and the property of the first Chern class $c_1(E\oplus F)=c_1(E)+c_1(F)$ one sees that $\Omega_i$'s are not independent but rather satisfy a relation
\bea\label{triv2}
\sum\limits_{i=1}^N\, \Omega_i = 0
\eea
The 2-forms $\Omega_i,\;i=1 \,\cdots\, N$, modulo the relation (\ref{triv2}), generate $\Hh^2(\fl_N, \Zz)$.

\vspace{0.5cm}

There is another interesting take on the relation (\ref{triv2}). It is related to Lagrangian submanifolds, or Lagrangian embeddings, which are a leitmotif of the present paper, and we feel it is time to introduce our main hero. What we want is a description of $\fl_N$, in which the forms $\Omega_i$ arise naturally. The first thing to appreciate in this direction is the existence of an embedding
\bea\label{embed}
i:\;\;\fl_N \,\hookrightarrow \, \underbrace{\CP^{N-1}\,\times\,\cdots\,\times \, \CP^{N-1}}_{N\;\textrm{times}} .
\eea
A point $m\in (\CP^{N-1})^{\times N}$ is a set of $N$ lines through the origin in $\CC^N$. Those points that correspond to $N$ \emph{orthogonal} lines are points of $\fl_N$ --- for this one should recall that $\fl_N$ may be thought of as a space of $N$ ordered orthogonal lines in $\CC^N$. Let us consider the line bundle $\mathcal{O}(1)_i$ over each $\CP^{N-1}_i$ factor. Then $\omega_i \sim c_1(\mathcal{O}(1)_i)$\;\footnote{$\sim$ means `in the same cohomology class'} can be taken as the Fubini-Study form on $\CP^{N-1}$. The forms $\Omega_i$ introduced above can be built simply as pull-backs of $\omega_i$ to $\fl_N$:
\bea\label{pullb}
\Omega_i =i^\ast(\omega_i)
\eea
With these ideas at hand, let us view $(\CP^{N-1})^{\times N}$ as a symplectic manifold with symplectic form
\bea\label{sfprod}
\omega=\sum\limits_{i=1}^N\;\omega_i
\eea
Our statement is that the embedding (\ref{embed}) is Lagrangian with respect to this symplectic form, i.e.
\bea\label{lagr1}
\omega|_{\fl_N}=0 .
\eea
We postpone the proof to Section \ref{momflag}. Taking into account (\ref{pullb}) and (\ref{lagr1}), the relation (\ref{triv2}) follows momentarily.

\vspace{0.3cm}
\noindent\;\;$\LHD$\;\; We have related the triviality of a certain line bundle over $\fl_N$ ($\overset{N}{\underset{i=1}{\oplus}}\;L_i = \fl_N\, \times \CC^N\;$) to the fact that $\fl_N$ is a Lagrangian submanifold of $(\CP^{N-1})^N$. \;\;$\RHD$ 

\vspace{0.3cm}

Now we are in a position to formulate the BWB result. To recall the setup, we are dealing with a representation of $U(N)$ with highest weight $\bl$, and in the sequel we will consider $\bl=(\lambda_1, \cdots, \lambda_N)$ to be the highest weight of the maximal torus $U(1)^N \subset U(N)$. The numbers $\lambda_i$ are integers. Construct the following line bundle on $(\CP^{N-1})^N$:
\bea
\tilde{L}_{\bl}=\mathcal{O}_1(\lambda_1)\otimes \cdots \otimes \mathcal{O}_N(\lambda_N)
\eea
Pulling it back to $\fl_N$, we get the line bundle of the BWB theorem:
\bea
L_{\bl}=i^\ast (\tilde{L}_{\bl})
\eea
The first Chern class of this bundle is equal to the following:
\bea\label{chernL}
c_1(L_{\bl})=i^\ast (\tilde{L}_{\bl}) =i^\ast (\sum\limits_{i=1}^N\;\lambda_i \,\omega_i)=
\sum\limits_{i=1}^N\;\lambda_i \,\Omega_i \equiv \Omega_{\bl}
\eea
The reason why we have been discussing this is that the pre-image of $c_1(L_{\bl})$ under the action of the external derivative $d$, i.e. the current $\{J:\,dJ=c_1(L_{\bl})\}$, is in the same cohomology class with the kinetic term in the path integral.

Any representation may be built on the sections of a line bundle over $\fl_N$, however for certain representations the base of the bundle may be reduced to a smaller space, i.e. a flag manifold of the form $\fl_{n_1, ... , n_m}=U(N)/U(n_1)\times\cdots \times U(n_m)$ with $\sum\limits_{j=1}^m\, n_j=N$ and not all $n_j$ equal to $1$. Notice that there is a fiber bundle $\pi: \fl_N \to \fl_{n_1, ..., n_m}$ --- this is in fact the first time that we encounter a so-called ``forgetful'' bundle, which will be used in Section \ref{forgetful}. The reduction of the base happens when the bundle $L_{\bl}$ over $\fl_N$ is the pull-back under $\pi$ of a fiber bundle over $\fl_{n_1, ..., n_m}$, i.e. when there exists a commutative diagram

\vspace{0.1cm}
\begin{center}
\hspace{2cm}
\begin{tabular}{ m{6cm} m{5cm} }
\quad
\xymatrix{
L_{\bl} \ar[r]\ar[d]_{\pi^\ast} & {\fl_N} \ar[d]^\pi \\
\tilde{L}_{\bl} \ar[r]& {\fl_{n_1, ..., n_m}}}
\end{tabular}
\end{center}

\vspace{0.1cm}

Not to get lost in the details and generalities, let us consider a simple example.

\textbf{Example.} There exists a fiber bundle $\fl_3 \to \CP^2$ with fiber $\CP^1$. One can take a bundle $\mathcal{O}(m) $ over $\CP^2$ (for positive $m$) and pull it back to $\fl_3$ --- in this way one obtains a representation of $U(3)$ which is a symmetric tensor power of degree $m$ of the fundamental representation.

The Lie algebraic description of what happens in this example and indeed generally is that the base space can be reduced when the highest weight $\bl$ is orthogonal to some of the roots $\alpha_i$ of $\mathfrak{su}_N$.

In order to clarify our notational conventions, we deduce the highest weight vector of the adjoint representation of $\mathfrak{su}_N$ by a straightforward calculation:\par\vspace{-0.5cm}
{\scriptsize
\bear\nonumber
\begin{pmatrix}
  v_1 & \0 & \cdots & \0 \\
  \0 & v_2 & \cdots & \0 \\
  \vdots  & \vdots  & \ddots & \0  \\
  \0 & \0 & \cdots & v_N
 \end{pmatrix}
 \cdot
\begin{pmatrix}
  \0 & \0 & \cdots & 1 \\
  \0 & \0 & \cdots & \0 \\
  \vdots  & \vdots  & \ddots & \0  \\
  \0 & \0 & \cdots & \0
 \end{pmatrix}
 -
 \begin{pmatrix}
  \0 & \0 & \cdots & 1 \\
  \0 & \0 & \cdots & \0 \\
  \vdots  & \vdots  & \ddots & \0  \\
  \0 & \0 & \cdots & \0
 \end{pmatrix}
 \cdot
 \begin{pmatrix}
  v_1 & \0 & \cdots & \0 \\
  \0 & v_2 & \cdots & \0 \\
  \vdots  & \vdots  & \ddots & \0  \\
  \0 & \0 & \cdots & v_N
 \end{pmatrix}
 =
 (v_1-v_N)\; \begin{pmatrix}
  \0 & \0 & \cdots & 1 \\
  \0 & \0 & \cdots & \0 \\
  \vdots  & \vdots  & \ddots & \0  \\
  \0 & \0 & \cdots & \0
 \end{pmatrix}
\eear}
\normalsize
Hence, the highest weight vector of the adjoint representation looks as follows:
\bea
{\boldsymbol \rho}=(1,0,...,0,-1)
\eea
The simple positive roots of $\mathfrak{su}_N$ can be found analogously and they have the following form in our notations:
\bea
{\boldsymbol \alpha}_1=(1,-1,0...0),\;{\boldsymbol \alpha}_2=(0,1,-1,...0),\;...\;,{\boldsymbol \alpha}_{N-1}=(0,0,0...,1,-1)
\eea

Therefore if for example $\bl\perp {\boldsymbol\alpha}_1$, it follows that $\lambda_1=\lambda_2$. It is convenient to portray this diagrammatically: if the highest weight $\bl$ is orthogonal to a root, one colors the node of the Dynkin diagram corresponding to this root \cite{FH}.

\textbf{Examples.}

The examples presented below are illustrating the general rule: to build the denominator of the coset $\fl_{n_1, ..., n_m}$, one assigns to each empty node a $U(1)$ factor and a $SU(M+1)$ factor to a group of $M$ adjacent colored nodes. The denominator of $U(N)/H$ is then $H=\prod\,\left(\textrm{these factors}\right)\,\times \,U(1)$.

\begin{tabular}{ c c }
  \begin{pspicture}(5,1)
\psset{linecolor=blue} 
\psset{linewidth=2pt} 
\cnode(1.5,0.5){.2}{b1}
\cnode[fillstyle=solid,fillcolor=yellow](3,0.5){.2}{b2}
\cnode[fillstyle=solid,fillcolor=yellow](4.5,0.5){.2}{b3}

\ncline{b1}{b2}
\ncline{b2}{b3}
\end{pspicture} & $\frac{U(4)}{U(3)\times U(1)}=\CP^3$ \\
  \begin{pspicture}(5,1)
\psset{linecolor=blue} 
\psset{linewidth=2pt} 
\cnode[fillstyle=solid,fillcolor=yellow](1.5,0.5){.2}{b1}
\cnode(3,0.5){.2}{b2}
\cnode[fillstyle=solid,fillcolor=yellow](4.5,0.5){.2}{b3}

\ncline{b1}{b2}
\ncline{b2}{b3}
\end{pspicture} & $\frac{U(4)}{U(2)\times U(2)}=G_2$ \\
  \begin{pspicture}(5,1)
\psset{linecolor=blue} 
\psset{linewidth=2pt} 
\cnode(1.5,0.5){.2}{b1}
\cnode[fillstyle=solid,fillcolor=yellow](3,0.5){.2}{b2}
\cnode(4.5,0.5){.2}{b3}

\ncline{b1}{b2}
\ncline{b2}{b3}
\end{pspicture}  & $\frac{U(4)}{U(2)\times U(1)\times U(1)}$\\
  \begin{pspicture}(8,1)
\psset{linecolor=blue} 
\psset{linewidth=2pt} 
\cnode(1.5,0.5){.2}{a1}
\cnode[fillstyle=solid,fillcolor=yellow](3,0.5){.2}{a2}
\cnode[fillstyle=solid,fillcolor=yellow](4.5,0.5){.2}{a3}
\cnode(6,0.5){.2}{a4}
\cnode[fillstyle=solid,fillcolor=yellow](7.5,0.5){.2}{a5}

\ncline{a1}{a2}
\ncline{a2}{a3}
\ncline{a3}{a4}
\ncline{a4}{a5}
\ncline{a5}{a6}
\ncline{a6}{a7}
\ncline{a7}{a8}
\end{pspicture} & $\frac{U(6)}{U(3)\times U(2)\times U(1)}$
\end{tabular}
\begin{center}
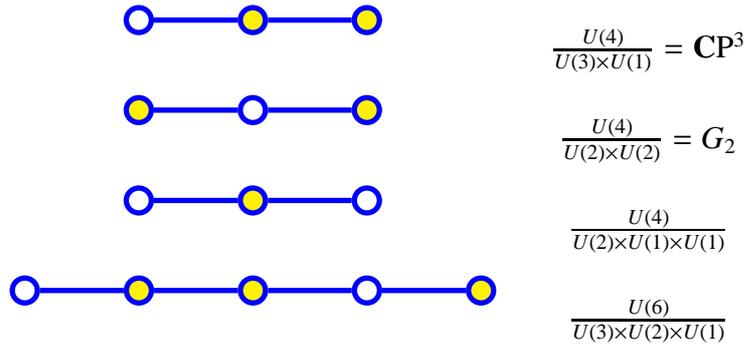
Fig. 2. Examples of coherent state manifolds.
\end{center}

\ssubsection{Coherent states. The BWB construction in practice.}\label{cohstates}

Coherent states are a type of basis in a vector space on which a Lie group $G$ is represented. One takes a highest weight vector $|v\rangle$ and forms its $G$-orbit, that is one considers all vectors of the form $g\,|v\rangle$, where $g\in G$. This is a continuous basis, which is therefore overcomplete, at least for a finite-dimensional representation. In what follows we will be dealing solely with the case of compact $G=U(N)$, however we find it useful to remind the reader of how the definition just introduced fits into the familiar setup of quantum mechanics. In this case one has a Heisenberg algebra $[a, a^\dagger]=\mathbbm{1}$ with a highest weight vector $|0\rangle$, which is annihilated by $a$ (and clearly fixed by the unit operator). The normalized coherent states are therefore given by the familiar formula
\bea
|v\rangle\equiv e^{-{1\over 2} |v|^2}\;e^{v\,a^\dagger}\,|0\rangle
\eea

\ssubsubsection{The basis.}

We start by building the bases in the relevant vector spaces using homogeneous polynomials. Let us take $SU(3)$ as a first example. In this case the polynomials will be built out of two sets of variables, $a_1, a_2, a_3$ and $b_1, b_2, b_3$. 

\textbf{a)} 
\hspace{0.3cm}\young(aaaa)\hspace{0.5cm}
Symmetric powers of the fundamental representation $\Rightarrow$ Symmetric polynomials in $a_1, a_2, a_3$ of degree $4$.

\textbf{b)} 
\hspace{0.3cm}
\young(aa,b) \hspace{0.5cm}
The adjoint representation $\Rightarrow$ Polynomials in $a, b$ of the form $a_i\,(a_j b_k-a_k b_j)$

\textbf{c)} 
\hspace{0.3cm}
\young(aaa,bb,c) \hspace{0.5cm}
The general $SU(N)$ case: 1) Assign to each row a letter $a, b, c, \cdots$. 2) For each column build antisymmetric combinations of the form $\sum\limits_{\sigma} \,(-)^\sigma\,a_{\sigma(i)}\,b_{\sigma(j)}\,c_{\sigma(k)}\,d_{\sigma(l)}$, where the number of letters participating is equal to the height of the column. 3) Multiply these antisymmetric combinations.

\textbf{Remark 1.} There are various linear relations among the polynomials built in the way described in \textbf{c)}. As a result, the representation is irreducible. For the example \textbf{b)} we could take the following polynomials as a basis
\bea\label{su3adj}
W^{ij}=a^i\,\epsilon^{jmn}\,a^m\,b^n ,\quad i, j=1 \cdots 3
\eea
There is a single relation $\sum\limits_{i=1}^3 W^{ii}=0$, therefore the dimension of the vector space is $3\times 3-1=8$, as it should be.


\textbf{Remark 2.} From this construction it follows that $b$ enters only in antisymmetric combinations with $a$, $c$ enters in antisymmetric combinations with $b$ and $a$, etc. Therefore the basis constructed above does not change under the transformation $b\to b+ r_1 a\, , \;c\to c+r_2 b+r_3 a$ for arbitrary $r_{1,2,3}$. 

\ssubsubsection{Coherent states from BWB.}

In the case of $SU(N)$ the coherent states are polynomials of a particular sort. Having the bases at hand, in order to build the coherent states all one needs to do is to pick a particular state and form its orbit under $SU(N)$. We will do it for the case of the three Young diagrams shown above, and the general case will be clear from these examples.

\textbf{a)} Highest weight vector $a_1^4$ leads to\;\;$\phi_v(a)=(\bar{v}\circ a)^4,\quad v\in \CP^{N-1}$

\textbf{b)} Highest weight vector $a_1\cdot(a_1 b_2-a_2 b_1)$ leads to\;\;$\phi_{uvw}(a, b)=(\bar{v}\circ a)\cdot[(\bar{v}\circ a)(\bar{w}\circ b)-(\bar{w}\circ a)(\bar{v}\circ b)],\quad \bar{w}\circ v=0$

\textbf{c)} Highest weight vector $a_1\cdot(a_1 b_2-a_2 b_1)\cdot(a_1 b_2 c_3-a_1 b_3 c_2-a_2 b_1 c_3-a_3 b_2 c_1+a_2 b_3 c_1+a_3 b_1 c_2 )$ leads to
\bear
&\phi_{uvw}(a, b, c)=(\bar{v}\circ a)\cdot[(\bar{v}\circ a)(\bar{w}\circ b)-(\bar{w}\circ a)(\bar{v}\circ b)]\cdot \\ \nonumber &\cdot[(\bar{v}\circ a)(\bar{w}\circ b)(\bar{u}\circ c)-(\bar{v}\circ a)(\bar{u}\circ b)(\bar{w}\circ c)-(\bar{w}\circ a)(\bar{v}\circ b)(\bar{u}\circ c)-\\ \nonumber
&-(\bar{u}\circ a)(\bar{w}\circ b)(\bar{v}\circ c)+(\bar{w}\circ a)(\bar{u}\circ b)(\bar{v}\circ c)+(\bar{u}\circ a)(\bar{v}\circ b)(\bar{w}\circ c)]
\eear
with $\bar{w}\circ v=\bar{u}\circ w=\bar{u}\circ v=0$.

In order to do calculations using these states one needs to know how to integrate over a flag manifold. Appendix \ref{intflag} provides an example --- a proof of the Parseval identity for the system \textbf{b)}.

\ssubsubsection{Relation to Schwinger-Wigner quantization.}

In brief, Schwinger-Wigner quantization is a way of representing spin operators using creation-annihilation operators (for a review see, for example, \cite{tsvelik}).

Suppose $\tau^\alpha$ are a set of $SU(N)$ generators in the fundamental representation. Introduce $N$ operators $a_i$ and their conjugates $a_i^\dagger$ with the canonical commutation relations
\bea\label{Weylalg}
[a_i, a_j^\dagger]=\delta_{ij}\, .
\eea
One can easily check that the operators
\bea
S^\alpha = a^\dagger_i \,\tau^\alpha_{ij}\,a_j,
\eea
satisfy the commutation relations of $\mathfrak{su}(N)$, and $S^\alpha$ act irreducibly on the subspace of the full Fock space specified by the condition
\bea
\sum\limits_{i=1}^N\,a_i^\dagger a_i = m,
\eea
where $m$ is a positive integer representing the `number of particles'. For a given $m$ the representation one obtains is the $m$-th symmetric power of the fundamental representation. What one should realize is that the $a_i$ are, morally speaking, the homogeneous coordinates $z_i$ on $\CP^{N-1}$. Indeed, if one imposes a partial gauge $\sum\,|z_i|^2=m$ in the path integral (\ref{evopxxx}), the kinetic term of the Lagrangian is simply $\mathcal{L}_0=\im\sum\,\bar{z}_i\circ \dot{z}_i$, therefore the canonical momentum $\pi_i=\frac{\dd \mathcal{L}_0}{\dd \dot{z}_i}=\im\bar{z}_i$, which leads to the algebra $\{z_i, \bar{z}_j\}=\delta_{ij}$, identical to (\ref{Weylalg}). Needless to say, this situation is general, and the correspondence holds for all representations.

To illustrate that this method is, nevertheless, not free from subtleties consider the $SU(3)$ adjoint representation of example \textbf{b)} from the previous section. To model this representation on a subspace of the Fock space we build the operators
\bear
N_1=a_1^\dagger a_1+a_2^\dagger a_2+a_3^\dagger a_3,\quad
N_2=b_1^\dagger b_1+b_2^\dagger b_2+b_3^\dagger b_3\\
O_1=a_1^\dagger b_1+a_2^\dagger b_2+a_3^\dagger b_3
\eear 
and we require the vectors $|\psi\rangle$ on which the representation is built to satisfy
\bea
N_1 |\psi\rangle = 2 \,|\psi\rangle,\;N_2 |\psi\rangle = |\psi\rangle,\;O_1 |\psi\rangle=0
\eea
The values of $N_1$ and $N_2$ correspond to the number of boxes in the first and second rows of the Young diagram. The expression for the $\mathfrak{su}_N$ generators looks as follows
\bea
S^\alpha=a_i^\dagger \tau^\alpha_{ij} a_j+b_i^\dagger \tau^\alpha_{ij} b_j\, ,
\eea
where $\tau^\alpha$ are the generators in the fundamental representation.

Notice that the classical condition $\bar{a}\circ b=0$ is translated to $O_1 |\psi\rangle=0$ with no counterpart $O_1^\dagger |\psi\rangle=0$. Indeed, the two equations would be incompatible, since $[O_1, O_1^\dagger]=N_1-N_2$ and $(N_1-N_2)\,|\psi\rangle=|\psi\rangle \neq 0$. One might worry that this introduces a certain asymmetry to the construction, however this asymmetry is the same one that is already present in the Young diagram. In the general case we should introduce $N$ creation operators $\mathbf{e}_k^\dagger$ for each row $k$ of the Young diagram ($k=1$ corresponds to the first row, i.e. the longest one), and impose the condition 
\bea
\mathcal{O}_{km}|\psi\rangle\equiv\mathbf{e}_k^\dagger\circ \mathbf{e}_m \;|\psi\rangle=0 \quad \textrm{for}\quad k<m
\eea
This is a compatible set of equations, since the operators $\mathcal{O}_{km}$ satisfy the algebra
\bea
[\mathcal{O}_{km}, \mathcal{O}_{np}]=\delta_{mn} \mathcal{O}_{kp}-\delta_{kp}\mathcal{O}_{nm}\quad\textrm{where}\quad k<m,\;\;n<p
\eea
$\mathcal{O}_{km}$ may be thus thought of as the positive roots of the Lie algebra $\mathfrak{su}_N$.

Apart from its aesthetic appeal, this construction offers certain calculational benefits, for instance the calculation of values of the Casimir operators on various representations becomes a matter of simple oscillator algebra (for an example see Appendix \ref{apposc}).

\ssubsection{The moment map for the action of loop rotations}\label{circaction}

In this Section we will look at the kinetic term in (\ref{actionxxx}) from a slightly different angle. As before, we will be assuming that $\MM$ is a symplectic manifold. Consider its loop space $\LM$, that is the space of all possible smooth embeddings of a circle $S^1$ into $\MM$. A point of the loop space $\gamma \in \LM$ is a loop $\gamma(t)\in \MM: \gamma(1)=\gamma(0)$. A tangent vector to $\LM$ at $\gamma$ is a periodic vector $\xi(t)\in T_{\gamma(t)}\MM: \xi(1)=\xi(0)$.

It is an important fact that one can, using the symplectic form $\Omega$ of $\MM$, define a symplectic form $\mathring{\Omega}$ on $\LM$. Indeed, suppose $\xi(t), \eta(t)$ are two tangent vectors to $\LM$ at $\gamma$. Then the symplectic form $\mathring{\Omega}$, evaluated on this pair of vectors, is:
\bea\label{sympl}
\mathring{\Omega}(\xi, \eta)=\int\limits_0^1\,dt\;\Omega(\xi(t),\eta(t))
\eea
Now notice that on $\LM$ there is an action of the group $S$ of shifts along the loop. Clearly, this group is isomorphic to $U(1)$, since loops are circles. In more detail, the action of a group element $g_\alpha \in S$ on a loop $\gamma(t)$ is given by
\bea
g_\alpha \circ \gamma(t)=\gamma(t+\alpha)
\eea
If we pick some local coordinates $x_i$ on $\MM$, then the vector field, which generates this action, can be written as follows\footnote{Note that the appearance of the functional derivative here is due to the fact that $\LM$ is an infinite-dimensional space. Despite this, the group $S$ is one-dimensional.}:
\bea
V=\int\,dt\;\dot{x}_i(t)\,\frac{\delta}{\delta x_i (t)}
\eea
This action also preserves the symplectic form (\ref{sympl}). What is the moment map associated with this action? In order to answer this question we evaluate $\mathring{\Omega}$ on $V$ to obtain a one-form:
\bea
\mathring{\Omega}(\bullet,V)=\int\limits_0^1\,dt\;\Omega(\delta\gamma(t),\dot{\gamma}(t))
\eea
or, in components,
\bea\label{symV}
\mathring{\Omega}(\bullet,V)=\int\limits_0^1\,dt\;\delta x^i(t)\, \Omega_{ij} \,\dot{x}^j(t)
\eea
We want to find such a function $\mu$ on $\LM$, whose variation under the contour change $\delta\gamma$ would produce the r.h.s. of (\ref{symV}). It turns out that such a function is nothing but the ``symplectic action''
\bea
\mu(\gamma)=\int_{D_\gamma}\;\Omega,
\eea
where $D_\gamma$ is a disc in $\MM$ having $\gamma$ as boundary. As it should, this expression, via the Stokes theorem, only depends on $\gamma$ --- the boundary of $D_\gamma$. The symplectic action is of course the same as the kinetic term in the classical action of the spin chain, for example the one in (\ref{actionxxx}).

\ssubsection{The action in supersymmetric form.}

Let $\mathcal{H}=\mathrm{Sym}(V_{\textrm{fund}}^{\otimes m })$ be the Hilbert space of a single spin. When the Hamiltonian is zero, $H=0$, the partition function of the spin may be written as follows (compare with \ref{actionxxx}):
\bea\label{dimH}
\mathcal{Z}= \tr(\mathbbm{1}) =\textrm{dim} \,\mathcal{H}= \int \; \prod\limits_{t\in[0,1]}\;d\mu(z(t),\;\bar{z}(t))\;\exp{\left(-m\;\int\limits_0^1\,dt\,   i\; \frac{\dot{z} \circ \bar{z}}{z\circ \bar{z}}\right)},
\eea
where $d\mu$ is the volume form on $\CP^{N-1}$.The volume form is proportional to the top power of the Fubini-Study form. Once we have picked some local real coordinates $x_1 ... \,x_{2N-2}$ on $\CP^{N-1}$, the Fubini-Study form may be written as $\omega=\omega_{ij}\, dx_i \wedge dx_j$. The volume form in turn can be expressed as $d\mu= \textrm{Pf}(\omega) \;dx_1 \wedge dx_2 \wedge ... \wedge dx_{2N}$, where $\textrm{Pf}(\omega)=\sqrt{\det{\omega}}$. On the other hand, there is an expression for the Pfaffian in terms of a Gaussian integral over real fermions: $\textrm{Pf}(\omega) = \int \,\; \prod\limits_i d\psi_i \; e^{\psi_i\;\omega_{ij}\; \psi_j}$. Using this observation (\ref{dimH}) may be rewritten as follows:
\bea
\mathcal{Z} = \int \; \prod\limits_{t\in[0,1]}\;dz_i(t) \wedge \bar{z}_i(t)\; \prod\limits_{t\in[0,1]}\,d\psi_i(t) \;\exp{\left(-m\;\int\limits_0^1\,dt\,   i\; \frac{\dot{z} \circ \bar{z}}{z\circ \bar{z}}+ \psi_i(t)\;\omega_{ij}(t)\; \psi_j(t) \right)}
\eea
The action in the exponent of this integral can be written in a manifestly supersymmetric form, i.e. in a sort of superspace. Indeed, introduce two complex conjugate fermionic coordinates $\theta, \bar{\theta}$ and the following `superfields':
\bear
&&Z_i(t, \bar{\theta}) = z_i(t)+{1\over \sqrt{m}}\,\bar{\theta} \, \psi_i(t),\quad\quad i=1 \ldots N\\
&&\bar{Z}_i(t, \theta) = \bar{z}_i(t)-{1\over \sqrt{m}}\,\theta \, \bar{\psi}_i(t)
\eear
Then there is a remarkably simple expression for the action:
\bear
&&\mathcal{S}=m\,\int\limits_0^1\;dt\;\int\;d\theta\;d\bar{\theta}\;\;\mathcal{K}(Z, \bar{Z}),\quad\textrm{where}\\
&&\mathcal{K}(Z, \bar{Z}) = \ln\left(\sum\limits_{i=1}^{N}\; Z_i \bar{Z}_i \right)
\eear
is the K\"ahler potential of $\CP^{N-1}$.

\ssection{The dynamical aspect.}\label{dynamical}

In the sequel we will be elaborating on Hamiltonians whose minima may be described as zero loci of moment maps. To this end we wish to remind the reader what the moment map is and recall its main properties.

\ssubsection{Properties of the moment map.}

Let $M$ a symplectic manifold with the symplectic form $\Omega$. Suppose there is an action of a Lie group $G$ on $M$ preserving the symplectic form, i.e. $\mathcal{L}_{X_a} \Omega =0$, where $X_a$ is a vector field on $M$ generating the action of a one-parametric subgroup of $G$ generated by the element $a\in \mathfrak{g}$ and $\mathcal{L}_{Y}=d\circ i_Y+i_Y \circ d$ is the Lie derivative.  Since $\Omega$ is closed by definition, $\mathcal{L}_{X_a} \Omega =0$ implies $d (i_{X_a}\Omega)=0$, therefore if $M$ is simply connected (it will be the case in all of the examples that we will consider), then $i_{X_a}\Omega = d \mu_a$, where $\mu_a$ is a function on $M$ and, of course, it can also be regarded as a function of $a$. In fact, since the vector field $X_a$ depends on $a$ linearly, $\mu_a$ is also a linear function of the Lie algebra element $a$, therefore, dropping the label $a$, i.e. considering all $a$'s at the same time, we may write that $\mu \in \mathfrak{g}^\ast$.

Let us summarize the above facts in the following definition: the moment map is a map $\mu: M \to \mathfrak{g}^\ast$ from a symplectic manifold $M$ to the dual of the Lie algebra $\mathfrak{g}$, possessing the following two properties:

(1) it is $G$-equivariant, i.e. $\mu(g\circ x)= Ad_g\,\mu(x)\equiv g\mu(x)g^{-1}$ for $x\in M, g\in G$.

(2) it is the generating function for Hamiltonians describing the action of $G$ on $M$, i.e. 
\bea\label{momham}
d\mu_a=i_{X_a}\Omega \quad \textrm{for} \quad a\in\mathfrak{g}.
\eea

We will mostly be dealing with a simple Lie group $G$. Its Lie algebra $\mathfrak{g}$ possess a unique $G$-invariant (Killing) scalar product, and therefore using this scalar product we will often forget the difference between $\mathfrak{g}$ and $\mathfrak{g}^\ast$.

\ssubsubsection{What if $\mu^{-1}(0)$ is a single orbit?}\label{orbtheo}

One important property of the moment map is that its zero-value set, usually denoted by $\mu^{-1}(0)$, is $G$-invariant, that is if $\mu(x)=0$ then $\mu(g\circ x)=0$: this is obvious from property (1). Therefore $\mu^{-1}(0)$ is a collection of $G$-orbits. Another fact, which will be cornerstone for the construction that follows, is that the restriction of $\Omega$ to each $G$-orbit in $\mu^{-1}(0)$ vanishes, i.e. each $G$-orbit in $\mu^{-1}(0)$ is an isotropic submanifold of $M$. This follows from (\ref{momham}) upon contraction with the vector field $X_b$ corresponding to a Lie algebra element $b$:
\bea
i_{X_b} d\mu_a \equiv \partial_b \mu_a = i_{X_b} i_{X_a} \Omega= \Omega(X_b, X_a)
\eea
The left hand side is zero, since $\partial_b \mu_a$ is the derivative of $\mu_a$ \textit{along} $\mu^{-1}(0)$. Therefore $\Omega(X_b, X_a)=0$, which means that the symplectic form is zero on vectors tangent to the orbit of $G$.

An isotropic submanifold $N\subset M$ can in principle have any dimension up to (and inclusive of) ${\mathrm{dim}\, M \over 2}$ --- in the latter case $N$ is called Lagrangian. There is a theorem which explains in what case a $G$-orbit in $\mu^{-1}(0)$ is Lagrangian: it is precisely when $\mu^{-1}(0)$ consists of one $G$-orbit, or in other words when  $\mu^{-1}(0)$ itself is a $G$-orbit\footnote{For a proof different from the one presented here see \cite{audin}.} \footnote{This is only true with the condition that there are $D-d$ linearly independent forms among $d\mu_a\big|_{\mu=0}$, where $d=\mathrm{dim} \,\mu^{-1}(0)$. Another way to put it is that the Jacobian $J\equiv{D \mu_a \over D x^i}\big|_{\mu=0}$ has rank $D-d$. This is a nondegeneracy condition, as can be seen from the following example: $M=\mathbb{R}^2=\{p, q\},\;\;G=SO(2),\;\;\mu=p^2+q^2 \Rightarrow \mathrm{dim}\,\mu^{-1}(0)=0,\;\;\mathrm{rank}\,J=0$. Here $\mu^{-1}(0)$ is a trivial orbit consisting of one point, but it is certainly not a Lagrangian submanifold. The above requirement means, in plain language, that the tangent vectors to $\mu^{-1}(0)$ are exactly those that annihilate the equation $\mu=0$ (i.e. they span the kernel of $d\mu_a$, which is hence $d$-dimensional).}. Indeed, assume that $\mu^{-1}(0)$ is a $G$-orbit. Therefore its tangent space is spanned by the vectors $X_a, a\in \mathfrak{g}$ introduced above. In general not all of them are linearly independent, so we pick a basis $X_1, ..., X_d$ of linearly independent vectors ($d=\textrm{dim}\; \mu^{-1}(0)$). We can assign to it $d$ one-forms: $\lambda_k = \Omega(\bullet, X_k)$. Since $X_k$ are linearly independent, $\lambda_k$ are linearly independent as well (since the form $\Omega$ is nondegenerate). Therefore the $d\times D \;(D=\textrm{dim} \,M > d=\textrm{dim} \,\mu^{-1}(0))$ matrix
\bea
\lambda=\{\lambda_1,\;\lambda_2, \,\hdots,\,; \lambda_d\}^T
\eea
has rank $d$. On the other hand, if $v$ is a null-vector of $\lambda$, it means that $\Omega(v, X_k)=0=\partial_v \mu_{a_k}$ for all $k$. This means that the equality $\mu=0$ is preserved along vector $v$, therefore $v$ is tangent to $\mu^{-1}(0)$ and is therefore expressed as a linear combination of the $X_k$. Since there are $d$ linearly independent vectors $X_k$, the nullity of $\lambda$ is $d$: $\textrm{null}\,\lambda = d$. By the rank-nullity theorem
\bea
\textrm{rank}\, \lambda + \textrm{null}\, \lambda = D \quad \Rightarrow \quad 2d=D,
\eea
which means that $\mu^{-1}(0)$ is Lagrangian.

The converse is also true, essentially by the same argument. Suppose $L=\mu^{-1}(0)$ is Lagrangian. Since a generic Hamiltonian vector for the Hamiltonian action of the group $G$ has the form $w_a=\Omega^{ij}\,\dd_j \mu_a\,{\dd \over \dd x^i}$ for some $a$, we need to show that there is a sufficient number of such independent vectors, more exactly ${D\over 2}$. Since $\Omega$ is nondegenerate, this is equivalent to showing that the matrix $\lambda=\{\Omega(\bullet, X_1),\,\cdots\,, \Omega(\bullet, X_{\mathrm{dim}\,\mathfrak{g}})\}^T$ has rank ${D\over 2}$. Similarly to what we had before, the kernel of this matrix is composed of those vectors $u$ that leave the moment map unchanged and equal to zero: $\dd_u \mu=0$. Such vectors are tangent to $L$, and therefore the nullity of $\lambda$ is equal to the dimension of $L$, i.e. ${D\over 2}$. The result follows once again from the rank-nullity theorem.  $\blacksquare$

\ssubsection{Moment maps for flag manifolds.}\label{momflag}

In this paper we are talking solely about manifolds of linear flags in complex vector spaces. Any such flag manifold is a quotient space (coset) $\fl(n_1, ... ,n_m)=U(N)/U(n_1)\times\cdots \times U(n_m)$. For the sake of practical calculations one usually writes a coset element as a $U(N)$-valued function $g(x)$ using some coordinates $x$. The action of $U(N)$, $x\to \tilde{x}$, is then presented as
\bea
g_0 \cdot g(x)= g(\tilde{x})\cdot h_0,\quad g_0\in U(N),\quad h_0 \in U(n_1)\times\cdots \times U(n_m)\, .
\eea
In order to write a moment map for this action we recall yet another way to think about flag manifolds. Every space $\fl(n_1, ... ,n_m)$ may be regarded as a (co)-adjoint orbit (adjoint and coadjoint representations are equivalent if there is a non-degenerate Killing metric, as it happens for $\mathfrak{su}_N$). It means that, as a model of $\fl(n_1, ... ,n_m)$, one can take an element $z\in \mathfrak{su}_N$ and consider its orbit $\textrm{Orb}(z)=\{g\cdot z\cdot g^{-1},\;\;g\in SU(N)\}$. One has to choose such $z$ that its stabilizer would be $U(n_1)\times\cdots \times U(n_m)$. In this case the moment map is simply
\bea
\mu(g)=g\cdot z \cdot g^{-1}
\eea
First of all, it has the right transformation property $\mu(g_0\cdot g)=g_0 \,\mu(g)\,g_0^{-1}$, which means that $\mu$ possesses property (1). To verify property (2) one needs to write the symplectic form on $\fl$ in terms of $z$ and $g$ . For this purpose we introduce the current $j=-g^{-1}\,\cdot\,dg$, which obeys the flatness (Maurer-Cartan) equation
\bea
dj-j\wedge j=0
\eea
In these terms the symplectic form is:
\bea
\Omega=\tr(z\, j\wedge j)
\eea
Due to the Maurer-Cartan equation, it is a closed form. We can assume that $z$ lies in the Cartan subalgebra, since clearly every orbit $\textrm{Orb}(z)$ intersects it. A simple calculation reveals that for $z$ in the Cartan subalgebra, $z=\textrm{diag}(\lambda_1,\, \cdots,\, \lambda_N)$, the form $\Omega$ coincides with $\Omega_{\vec{\lambda}}$ introduced in \ref{chernL}. Suppose now that $v_a$ is a vector field on $\fl$ corresponding to Lie algebra element $T_a$. Then one can verify that
\bea
j(v_a)\equiv i_{v_a}\, j =-g^{-1}\,\nabla_{v_a}g=-g^{-1}\frac{d}{dt}\left(e^{T_a\, t}\,g\right)\big|_{t=0}=- g^{-1} T_a g
\eea
Using this, it is straightforward to check the defining property (2) of the moment map:
\bea
d(\tr (\mu T_a))=d(\tr (g\,z\,g^{-1} \,T_a))=\tr(z\,[j,g^{-1}T_a g])=i_{v_a}\,\tr(z j\wedge j)=i_{v_a}\,\Omega
\eea

\ssubsection{The Hamiltonian.}\label{hamsec}

After this general discussion we come to the actual Hamiltonians. The Hamiltonians, which we will consider, are built from interactions of the form $\kappa_{mn}\,S^m_i S^n_{j}$, where $i, j$ are the sites of the spin chain, and $\kappa$ is the Killing form\footnote{Such interaction is more easily visualizable when written in the form $\vec{S}_i \cdot \vec{S}_{j}$ .}. We will assume that the spin chain is \emph{translationally invariant}, i.e. its Hamiltonian can be defined by shifting along the chain of a Hamiltonian of a `unit cell'. The number of sites in the unit cell will depend on the target space that we want to get in the sigma model. However, for unit cell of length $m$ the Hamiltonian is of the form\footnote{The Hamiltonian considered in \cite{Bykov}, $H=\sum\limits_{i=1}^{L} \, (P_{i,i+1}+{1\over 2}P_{i,i+2})$, is a particular case when $m=3$.}
\begin{empheq}[box=\fbox]{align}
\label{genham}
\hspace{3em}
&\mathcal{H}=\sum\limits_{i=1}^{L} \, \sum\limits_{k=1}^{m-1} \, d_k\; \vec{S}_i\cdot \vec{S}_{i+k},
\hspace{3em} \\ \nonumber
&\textrm{where} \\ \label{dk}
&d_k=\sqrt{\frac{m-k}{k}}
\end{empheq}
The expression for $d_k$ is derived in the Appendix \ref{A} --- it is the unique result if one insists on the two-dimensional Lorenz invariance of the resulting sigma model.

In order to write the Hamiltonian in terms of the coherent states we note that for expressions quadratic in the spins this can be done simply by replacing the spins $\vec{S}_i$ by the corresponding moment maps $\mu_i \in \mathfrak{su}_N$. A fully honest calculation would involve the construction of a path integral `from scratch' --- the interested reader is referred to \cite{Bykov} for an idea of how this can be done. In any case, the Hamiltonian has the following form, when written in coherent states:
\bea\label{hamcoh}
\mathcal{H}\to \sum\limits_{i=1}^{L} \, \sum\limits_{k=1}^l \, d_k\; \tr(\mu_i\,\mu_{i+k})
\eea
We will assume that the manifold of coherent states is the Grassmannian $G_n\equiv {U(N)\over U(n)\times U(N-n)}$ for some $n$. It will be explained in the next Section why we can restrict to this case. The moment map for the action of $SU(N)$ on a Grassmannian $G_n$ of $n$-planes has the form
\bea\label{momgrass}
\mu=\sum\limits_{k=1}^{n}\;\frac{\bar{z}_k\otimes z_k}{\bar{z}_k\circ z_k}-\frac{n}{N}\,\mathbbm{1},
\eea
where the vectors $\{ z_k\}$ form an orthogonal basis in a given $n$-plane: $\bar{z}_m\circ z_n=\delta_{m n} \, \bar{z}_m \circ z_m$. It is not difficult to see that the Hamiltonian (\ref{hamcoh}) is a sum of positive terms (apart from some irrelevant constants), and the minimum is attained when all the $z$ vectors at neighboring $m$ sites are orthogonal. This means that the corresponding $n_i$-dimensional planes ($i=1\cdots m$) are orthogonal to each other (and together fill the vector space $\CC^N$). This configuration is precisely what we mean by the \emph{classical antiferromagnetic vacuum}.


\ssubsection{General equivariant Lagrangian embeddings: \\\quad\quad \quad \;\,forgetful fiber bundles.}\label{forgetful}

In this section we discuss the geometric origins of the Lagrangian embeddings which we have built using the moment map in the previous sections. The question we want to answer is: how big is the class of flag manifolds $M, N$ such that there exist $G$-equivariant Lagrangian embeddings $M \underset{\textrm{Lagr}}{\hookrightarrow} N$ ? The answer that we will find is that for each flag manifold $M$ there is a canonical embedding into a product of symmetric spaces (Grassmannians) $N$.

\noindent\textbf{Example. $M=\mathcal{F}_3, N=(\CP^2)^{\times 3}$ .}

We start once again from our basic example (already discussed in Section \ref{BWB}), as it illustrates the general situation quite well. Recall that $\mathcal{F}_3$ is interpreted geometrically as a space of ordered 3-tuples of orthogonal three lines in $\mathbf{C}^3$. Therefore there exist three fiber bundles, which associate with a given 3-tuple $(v_1, v_2, v_3)$ one of the three lines, either $v_1$, $v_2$ or $v_3$:\newline
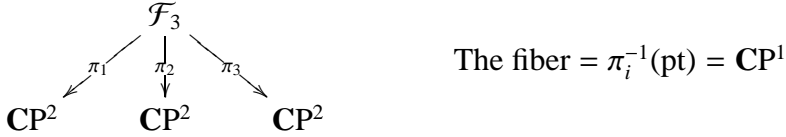
\begin{tabular}{ m{6cm} m{5cm} }
\quad
\xymatrix{
 & { \mathcal{F}_3 } \ar[dl]|{\pi_1} \ar[d]|{\pi_2} \ar[dr]|{\pi_3} & \\
{\CP^2} & {\CP^2} & {\CP^2}}
& 
The fiber $=\pi_i^{-1}(\textrm{pt})=\CP^1$
\end{tabular}
\newline
Since each of these fiber bundles `forgets' two lines out of three, they may be called `forgetful' fiber bundles. They are explicitly $G=SU(3)$-equivariant. The embedding under consideration is seen to be the map $M \to \pi_1(M)\,\times\,\pi_2(M)\,\times\,\pi_3(M) = N$. Since we know this map is injective, or in other words that it does not send any two distinct points $a, b\in M$ to the same one in $N$, let us discuss what it means geometrically. If it were not injective, that would mean that $a$ and $b$ lie simultaneously in all three fibers $f_1=\pi_1^{-1}(m)$, $f_2=\pi_2^{-1}(n)$ and $f_3=\pi_3^{-1}(p)$ of the corresponding fibrations, that is to say $(a,b)\in f_1\cap f_2 \cap f_3$. Therefore we come to the conclusion that any three fibers intersect in no more than one point. We can even be more specific: if $m, n, p$ are not mutually orthogonal, then the fibers of the corresponding fiber bundles do not intersect at all, whereas if they \textit{are} mutually orthogonal, then the intersection consists of one point.

We now wish to generalize the above example to the case of a general flag manifold
\bea
\mathcal{F}_{n_1,\,\cdots\, , n_m}= U(N)/U(n_1)\times \cdots \times U(n_m).
\eea
We can now build $m$ fiber bundles by forgetting the `fine structure' of the flag and remembering only one linear subspace (and its orthogonal) at a time:\newline
\vspace{-0.4cm}
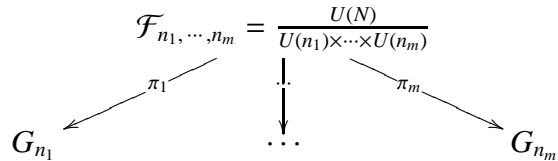
\begin{center}
\begin{tabular}{ m{10cm} m{5cm}}
\quad
\xymatrix{
 & { \mathcal{F}_{n_1,\,\cdots\, , n_m}= \frac{U(N)}{U(n_1)\times \cdots \times U(n_m)} } \ar[dl]|{\pi_1} \ar[d]|{\cdots} \ar[dr]|{\pi_m} & \\
{G_{n_1}} & {\cdots} & {G_{n_m}}
}
\end{tabular}
\end{center}
\vspace{0.1cm}
where $G_{n}={U(N) \over U(n)\times U(N-n)}$ is the Grassmannian of $n$-planes in $\CC^N$.

The corresponding map
\bea \label{embed2}
\mathcal{F}_{n_1,\,\cdots\, , n_m} \hookrightarrow \prod\limits_{j=1}^m\;G_{n_j}
\eea
is an embedding and, moreover, it is a Lagrangian embedding. First let us perform a dimensionality check:
\bear
\mathrm{dim}\,\mathcal{F}_{n_1,\,\cdots\, , n_m} = N^2-\sum\limits_{j=1}^m\,n_j^2,\quad \mathrm{dim}\,G_n=N^2-n^2-(N-n)^2=2(n\cdot N-n^2)\\
\Rightarrow \mathrm{dim}\, \prod\limits_{j=1}^m\;G_{n_j}=2\;\sum\limits_{n=1}^m\,(n\cdot N-n^2)=2\;\left(N^2-\sum\limits_{n=1}^m n^2 \right)=2\; \mathrm{dim}\,\mathcal{F}_{n_1,\,\cdots\, , n_m}
\eear
Given the background accummulated to this moment, it is not difficult to show that the embedding is Lagrangian. We need to construct a moment map for the diagonal action of $SU(N)$ on the product of Grassmannians and prove that $\mu^{-1}(0)$ is the flag manifold under consideration. We have in fact already constructed the moment map for a single Grassmannian in (\ref{momgrass}), so now we take a sum of those:
\bea
\mu=\sum\limits_{i=1}^m\;\sum\limits_{k=1}^{n_i}\;\frac{\bar{z}_{i_k}\otimes z_{i_k}}{\bar{z}_{i_k}\circ z_{i_k}}-\mathbbm{1}
\eea
where we have used the relation $\sum\limits_{i=1}^m\, n_i=N$. One should recall that in this formula it is implied that $\bar{z}_{i_m}\circ z_{i_n}=\delta_{mn}$. On the other hand, the set $\mu^{-1}(0)$ is composed of $N$-tuples of orthogonal $z$-vectors. It follows that the $z$-vectors representing different $n_i$-dimensional planes in $\CC^N$ ($i=1\cdots m$) are mutually orthogonal. The set of such orthogonal subspaces is precisely the flag manifold $\mathcal{F}_{n_1,\,\cdots\, , n_m}$\,!

\ssection{The metric.}\label{metricsec}


On a general flag manifold, which is not necessarily a symmetric space, there may exist a whole family of $SU(N)$-invariant metrics\footnote{For an example of metrics on the complete flag manifold $U(N)/U(1)^N$ see \cite{Bykov}.}. The construction of continuum limits that we are discussing in the present paper provides a particular representative from that family. The aim of the present section is to give an intrinsic and universal expression for the metric that arises in this context.

Let us recall the general setup, to which the remarks of the present section are generally applicable. One has a symplectic manifold $(\mathcal{M}, \omega)$, a function $I$ on $\mathcal{M}$ and a Lagrangian submanifold $L\subset \mathcal{M}$, on which $I$ has a minimum. We can form the Hessian of the function $I$:
\bea
h_{ij}=\frac{\dd^2 I}{\dd x^i \dd x^j}
\eea
It is worth noting that on the critical set the Hessian transforms as a tensor, i.e. under the change of coordinates $x\to y(x)$ we have $\frac{\dd I}{\dd y^i}=0$ and
\bea
\frac{\dd^2 I}{\dd y^i\,\dd y^j}=\frac{\dd}{\dd y^i}\left(\frac{\dd x^k}{\dd y^j}\cdot \frac{\dd I}{\dd x^k}\right)=\frac{\dd}{\dd y^i}\left(\frac{\dd x^k}{\dd y^j}\right)\cdot\underbrace{\frac{\dd I}{\dd x^k}}_{=0\,\textrm{on}\, L}+\frac{\dd x^m}{\dd y^i}\,\frac{\dd x^n}{\dd y^j}\,\frac{\dd^2 I}{\dd x^m\dd x^n}
\eea
Besides, since on $L$ one has $\frac{\dd I}{\dd x^i}=0$, for any vector $v\in T_p L$ tangent to $L$ we have $\nabla_v \frac{\dd I}{\dd x^i}=0=h_{ij}\,v^j$, therefore $v$ is a zero-vector of $h$. An extra requirement that we will impose on the system and that is always fulfilled in our applications is that $h$ is non-degenerate when restricted to the vectors normal to $L$\footnote{This is sometimes called `nondegeneracy in the sense of Bott'.}. Because $L$ is a minimum for $I$, $h$ is non-negative-definite. The latter two statements can be summarized by saying that $h$ is the metric on the normal bundle to $L$ in $\mathcal{M}$. What we want, however, is a metric on the tangent bundle to $L$. One of the properties that makes a Lagrangian submanifold special and different from a generic submanifold is that its tangent bundle is isomorphic to its conormal bundle. The isomorphism is, in fact, provided simply by the symplectic form $\omega$: indeed, let $v$ be a tangent vector to $L$, then $\lambda=i_v\omega$ is a one-form on $\mathcal{M}$. Its kernel is composed of those vectors that are tangent to $L$, so $\lambda$ may be viewed as a one-form on $\mathcal{N}L\subset T\mathcal{M}$, i.e. $\lambda \in \mathcal{N}^\ast L$. The metric on $\mathcal{N}L$ that we have just constructed provides an isomorphism $\mathcal{N}L\simeq \mathcal{N}^\ast L$, hence we obtain a metric on $L$ of the form
\begin{empheq}[box=\fbox]{align}
\label{metric1}
\hspace{2em}
g_{ij}=\omega_{im}\cdot \left[\left(\frac{\dd^2 I}{\dd x^2}\right)^{-1}\right]^{mn}\cdot \omega_{nj}=\omega_{im}\,h^{mn}\,\omega_{nj}\hspace{2em}
\end{empheq}
In this formula it is implied that the Hessian $\frac{\dd^2 I}{\dd x^2}$ has to be restricted first to the directions normal to $L$ and only then it can be inverted. $g_{ij}$ is positive definite, since $h$ is positive-definite: $v^i g_{ij} v^j=u_i h^{ij} u_j \geqslant 0\quad (u_i=\omega_{ij}v^j)$. Therefore $g_{ij}$ is the metric that we were looking for. 


Notice that the expression for the metric (\ref{metric1}) is valid in the quite general setup outlined at the start of this Section. The metrics on flag manifolds that are obtained from a spin chain by means of a continuum limit  represent particular applications of this construction. As it follows from the formulas in Appendix \ref{A}, in particular from (\ref{fulllagr}), in our case the function $I$ is as follows (here $\mu_i$ is the moment map on the $i$-th Grassmannian):
\begin{empheq}[box=\fbox]{align}
\label{Ifunc}
\hspace{1em}
I=\sum\limits_{1=i<j}^{m}\,(d_{j-i}+d_{m-(j-i)})\,\tr(\mu_i\,\mu_j)=m\;\sum\limits_{1=i<j}^{j=m}\,\frac{\tr(\mu_i\,\mu_j)}{\sqrt{(j-i)\,(m-(j-i))}}\hspace{0.5em}
\end{empheq}

\ssection{The topological term.}\label{topsec}

Similarly to the case of the metric, one can build a universal and transparent expression for the topological term that arises in the continuum limit of a spin chain. The idea is that one can, once again, exploit the fact that there exists an embedding of the type (\ref{embed2}) of a general flag manifold $\mathcal{F}_{n_1,\,\cdots\, , n_m}= U(N)/U(n_1)\times \cdots \times U(n_m)$ into a product of symmetric spaces:
\bea\label{embedprod}
i:\,\mathcal{F}_{n_1,\,\cdots\, , n_m} \,\hookrightarrow\, G_{n_1}\times ... \times G_{n_m},
\eea
where $G_n=U(N)/U(n)\times U(N-n)$ is a Grassmannian (symmetric space). The most natural fiber bundle to construct over such Grassmannian is the tautological bundle of $n$-planes that we will denote by $\mathcal{O}_{G_n}(1)$, analogously to the case of $\CP^{N-1}$. The cohomology ring of the original flag manifold $\mathcal{F}_{n_1,\,\cdots\, , n_m}$ may be built as a pull-back of the corresponding cohomology ring of the Grassmannians. In particular, let $r_k=i^\ast\big( c_1(\mathcal{O}_{G_{n_k}}(1)) \big)$ be the (pull-back of the) first Chern class of the tautological bundle. $r_k$ satisfy a single relation
\bea\label{relr}
\sum_{k=1}^m\,r_k=0
\eea
Then any element of $\Hh^2(\mathcal{F}_{n_1,\,\cdots\, , n_m}, \Zz)$ can be written as a linear combination with integer coefficients of the pull-backs of these Chern classes. Finding the integer coefficients that describe the 2-form $\Omega$ arising in the $\theta$-term is the goal of this Section.

The construction is, in fact, rather elementary. We start from a general expression for the topological term:
\bea
m\,\Omega=\sum\limits_{k=1}^m\, a_k\, r_k \quad \mathrm{mod}\;\;m,
\eea
where $a_k$ are integers. The crucial requirement, which follows from the translational invariance of the original Hamiltonian (\ref{genham}), is that $\Omega$ should be invariant under a cyclic permutation of the positions of the Grassmannians (indeed, their cyclic position simply the way we `cut' the spin chain into elementary cells, and this should not affect the result). Denoting the permutation by $\Pi$, we can formalize this requirement in the following way:
\bea\label{perm1}
\Pi(m\, \Omega)= \sum\limits_{k=1}^m\, a_{k+1}\, r_k = m\,\Omega\quad \mathrm{mod}\;\;m,
\eea
When dealing with this equation, one should recall that there is a relation (\ref{relr}) on the $r$'s, therefore (\ref{perm1}) may be rewritten as
\bea
\sum\limits_{k=1}^m\, a_{k+1}\, r_k =\sum\limits_{k=1}^m\, a_{k}\, r_k + n\,\sum\limits_{k=1}^m\, r_k \quad \mathrm{mod}\;\;m\, ,\quad n\in \mathbf{Z}
\eea
or in other words
\bea
a_{k+1}-a_k=n\quad \mathrm{mod}\;\;m
\eea
Therefore $a_k = k \cdot n\; \mathrm{mod}\;\;m$. In fact, $n$ can be adjusted at will by taking symmetric powers of representations at all nodes, so for the minimal choice $n=1$ the $\theta$-term can be written as
\begin{empheq}[box=\fbox]{align}
\label{theta1}
\hspace{1em}
\Omega=\frac{1}{m}\,\left(\sum\limits_{k=1}^m\, k\,\cdot \, r_k \right)
\hspace{1em}
\end{empheq}
As it should, the topological term of this form does not depend on the \emph{cyclic} ordering of the flag manifolds. Nevertheless, it follows from (\ref{theta1}) that it certainly does depend on their ordering (up to cyclic permutation). Therefore permuting the sites of the spin chain, putting the representations in a different order, changes the topological term.

In particular, we come to the following interesting conclusion:

\vspace{0.3cm}
\noindent\;\;$\LHD$\;\; The basis of the cohomology group $\Hh^2(\mathcal{F}_{n_1,\cdots, n_m}, \mathbf{Z}_m)$ can be obtained by permuting the sites of the spin chain.\;\;$\RHD$

\vspace{0.3cm}

It is seen from (\ref{theta1}) that the value of $\theta$ is $\theta={2\pi \over m}$. It also follows from the general discussion of Section \ref{pathint} that if one replaces the original representations at each site of the spin chain by their symmetric tensor products of degree $r$, the value of $\theta$ is multiplied by $r$ as a result.

\ssection{Discussion.}\label{discuss}

In the present paper we had a two-fold goal: to give an overview of the geometrical approach to representation theory (the Borel-Weil-Bott theorem) and coherent states and, using them, to formulate two results concerning the long-wavelength limits of certain spin chains. These infrared limits are two-dimensional sigma models, whose target space can be an arbitrary flag manifold (though in the present paper we restrict ourselves to the case of flags in complex vector spaces, i.e. the symmetry group $U(N)$). We have shown that for a flag manifold of our wish a spin chain can be built with a continuum limit described by the sigma model with this flag manifold as its target space. The Hamiltonian of this spin chain is given by (\ref{genham}), (\ref{dk}). From a mathematical point of view, our construction relies on two facts:

1) The Hamiltonian is a function on a product of Grassmannians, which has a minimum on a Lagrangian submanifold.

2) There exists a Lagrangian embedding of any flag manifold into a product of Grassmannians (\ref{embedprod}).

The submanifold, on which the Hamiltonian reaches a minimum, may be viewed as the quasiclassical antiferromagnetic vacuum. It has to be Lagrangian, since it is only in the vicinity of a Lagrangian submanifold $L$ that the original symplectic manifold (the product of Grassmannians) looks as the cotangent bundle to $L$ --- the phase space of the sigma model that we are building. From the more technical point of view, it is precisely this circumstance that allows us to integrate over the momenta $p$, cotangent to $L$ (see Fig. 1), and obtain an action quadratic in time derivatives (rather than linear in them, like the original action).

The meaning of the second requirement is the following. The representations sitting at the sites of the spin chain that we are considering are the ones appearing as spaces of sections of holomorphic fiber bundles over Grassmannians. Therefore a product of Grassmannians represents the union of several consecutive sites of the spin chain (the elementary cell). The Hamiltonian, which is a function on this product of Grassmannians, is then extended to the full spin chain by translational invariance. The fact that \emph{any} flag manifold can be embedded as a Lagrangian submanifold into a product of Grassmannians simply means that for a given flag manifold we can always find a spin chain realizing its geometry in the continuum limit.

The two main results of the paper are given by formulas (\ref{metric1}) and (\ref{theta1}). They provide rather explicit expressions for the metric and topological term of the resulting sigma models. We have found that in the situation when a function $I$ on a symplectic manifold $M$ has a non-degenerate\footnote{In the sense of Bott} minimum on a Lagrangian submanifold $L\subset M$, there is a canonical metric on $L$ that can be built using this data. It is given by formula (\ref{metric1}). The function $I$ in our case is given by \ref{Ifunc}.

As we have discussed in Section \ref{topsec}, the topological term can be obtained by a very simple procedure. We form a linear combination of the first Chern classes of plane (tautological) bundles over the Grassmannians into which our flag manifold is embedded (at this point it is crucial to choose an ordering of the products) and then demand its invariance under cyclic permutation of the Grassmannians in the product. This is a natural requirement, since the cyclic permutation corresponds to a shift along the spin chain (or, equivalently, a different partition of the spin chain into elementary cells), which should not change the final result. There are two remarkable facts about the result (\ref{topsec}). The first one is that what enters the denominator in (\ref{topsec}) is $m$ --- the number of Grassmannians in the product, or the number of $U(n_i)$ factors in the denominator of a fraction which describes the flag manifold as a homogeneous space: $\frac{U(N)}{U(n_1)\times \cdots \times U(n_m)}$. This means that the $\theta$-term belongs to the second cohomology group of the flag manifold with coefficients in $\mathbf{Z}_m$. It would be interesting to understand if this implies any $\textrm{mod}\; m$ periodicity of the mass gap in the spin chain. Another thing to notice is that in order to determine the $\theta$-term we have chosen an ordering of the Grassmannians in the product, or in other words the ordering of sites in the spin chain. Therefore a different ordering gives a different $\theta$-term. One can generate the cohomology group $\Hh^2(\mathcal{F}, \mathbf{Z}_m)$ by permuting the sites of the spin chain!

\comment{
The first Chern class of the flag manifold $c_1(\mathcal{F}_3)$ can be calculated in the following way. The main idea is that the holomorphic tangent bundle $T_h(\mathcal{F}_3)$ can be expressed in terms of the three line bundles over $\mathcal{F}_3$ defined in the previous section and which we will denote by $L_1, L_2, L_3$. The sum of these three bundles is a trivial vector bundle: $L_1 \oplus L_2 \oplus L_3 = \mathcal{F}_3 \times \mathbb{C}^3$. Therefore we get a relation for the first Chern classes $c_{(1)}, c_{(2)}, c_{(3)}$ of these bundles:
\bea\label{relchern}
(1+c_{(1)})(1+c_{(2)})(1+c_{(3)})=1
\eea
Then the holomorphic tangent bundle of the flag manifold may be written in the form
\bea
T_h(\mathcal{F}_3) = \sum\limits_{i<j=1}^3\; L_i \otimes \bar{L}_j
\eea
Since from (\ref{relchern}) it follows that $c_{(1)} + c_{(2)} + c_{(3)}=0$, we can choose to eliminate $c_{(2)}$ from all formulas, which leads to
the answer
\bea
c_1(T_h(\mathcal{F}_3)) = 2 (c_{(1)}-c_{(3)})
\eea
}


\begin{center}
{\normalfont\scshape \large Acknowledgments}
\end{center}

I am grateful to Profs. S.Frolov, K.Zarembo for discussions. I am grateful to Prof. E.Witten for his remarks to my talk at the conference ``MathPhyz 2011''. I am especially indebted to Prof. A.A.Slavnov for constant support and encouragement. My work was supported in part by grants RFBR 11-01-00296-a, 11-01-12037-ofi-m-2011
and in part by grant for the Support of Leading Scientific Schools of Russia NSh-4612.2012.1.
\vspace{0.7cm}
\appendix

\begin{center}
{\normalfont\scshape\Large\underline{Appendices}}
\end{center}
\vspace{-0.7cm}
\ssection{Derivation of the metric and the $\theta$-term.}\label{A}

In this Appendix we perform a complete calculation, which is rather similar to the one of \cite{Bykov}, but more general. We have shown before that a generic flag manifold may be embedded in a product of Grassmannians. Therefore we can restrict to the case when the representations can be built from sections of fiber bundles over Grassmannians. We adopt the simplest possible model for a Grassmannian. Consider the case of 
\bea
G_n=\frac{U(N)}{U(n)\times U(N-n)}, 
\eea
We will represent it with $N$ orthonormal complex vectors $u_k$, $\bar{u}_m \circ u_n=\delta_{mn}$, where equivalence relations are imposed on the sets of lines $\{u_1, \cdots,u_{n} \}, \{u_{n+1},\cdots, u_{N}\}$. Each such set represents a plane --- the linear span of the corresponding vectors --- therefore, for example, $\{u_1, \cdots, u_{n}\} \sim \{u'_1, \cdots, u'_{n}\}$, if the two sets are related by a $U(n)$ rotation of the basis. The vectors $u_{1}, \cdots, u_{n}$ will enter all calculations only in $U(n)$-invariant combinations. Call $\omega$ the symplectic form on $G_n$. The current $J$ defined by $dJ=\omega$ can be built in the following way
\bea
J=\im \, \sum\limits_{i=1}^n\,\bar{u}_i \circ \dot{u}_i
\eea
This is invariant with respect to the $SU(n)$ gauge transformations $u_i \to g_{ij}\,u_j,\;\;g\in SU(n)$, since
\bea
J\to J+\im \underbrace{\bar{u}_i \circ u_k}_{=\delta_{ik}} \;(g^\dagger \dot{g})_{ik}=J,\quad\textrm{since}\;\tr(g^\dagger\,\dot{g})=0
\eea
(For the case $g\in U(1)$ a total derivative is added to $J$, and therefore the integrated $J$-current is invariant).

We will start from a Hamiltonian of the following general form:
\bea\label{hamgen2}
\mathcal{H}=\sum\limits_k\,\sum\limits_{s=1}^{m-1}\,d_s\, \vec{S}_k\cdot \vec{S}_{k+s}\;.
\eea
We therefore restrict to interactions of range $m-1$. Any consecutive $m$ sites will be therefore called a `unit cell', or `elementary cell'.

Suppose the elementary cell is built of $l$ sites with a Grassmannian $G_{n_i}$ sitting at the $i$-th site ($i=1\cdots l$). One builds a moment map
\bea
\mu_i=\sum\limits_{k=1}^{n_i}\;\bar{u}_k\otimes u_k-\frac{n_i}{N}\,\mathbf{1} .
\eea
Then the spin-spin interaction of the form $\vec{S}_i\cdot \vec{S}_j$ leads to the term
\bea
\tr(\mu_i\,\mu_j) \sim 2\,\sum\limits_{m<n} \,|\bar{u}_m^{(i)}\circ u_n^{(j)}|^2 + \textrm{constant terms}
\eea
 in the coherent state Hamiltonian. Therefore essentially the only difference from the case of $\CP^{N-1}=G_1$ considered in detail in \cite{Bykov} is that now we have to sum over several similar interaction terms. It is convenient to depict diagrammatically a site with Grassmannian $G_n$ as a collection of $n$ points aligned vertically.
\vspace{0.2cm}\newline
\begin{center}
 \begin{pspicture}(10,2)
\psset{linecolor=blue} 
\psset{linewidth=1pt} 
\psset{fillcolor=blue} 
\psset{linestyle=dashed}
\psframe[linewidth=2pt,framearc=.3,fillstyle=solid, fillcolor=yellow](1.5,0.8)(5.5,2.2)
\pnode(0,1.5){ap0}
\cnode*(1,1.5){.1}{ap1}
\cnode*(2,1.5){.1}{ap2}
\cnode*(3,1.25){.1}{ap3}
\cnode*(3,1.75){.1}{ap4}
\cnode*(4,1.25){.1}{ap5}
\cnode*(4,1.75){.1}{ap6}
\cnode*(5,1.5){.1}{ap7}
\cnode*(6,1.5){.1}{ap8}
\cnode*(7,1.25){.1}{ap9}
\cnode*(7,1.75){.1}{ap10}
\cnode*(8,1.25){.1}{ap11}
\cnode*(8,1.75){.1}{ap12}
\cnode*(9,1.5){.1}{ap13}
\pnode(10,1.5){ap14}

\rput(2.1,1.75){\footnotesize  $u_{1_1}$}
\rput(3.35,1.75){\footnotesize  $u_{2_1}$}
\rput(3.35,1.25){\footnotesize  $u_{2_2}$}
\rput(4.35,1.75){\footnotesize  $u_{3_1}$}
\rput(4.35,1.25){\footnotesize  $u_{3_2}$}
\rput(5.1,1.75){\footnotesize  $u_{4_1}$}

\pscurve[linecolor=red,linestyle=solid]{->}(7,2.5)(6.2,2.4)(5.4,2.15)
\rput(8.2,2.5){\footnotesize  Elementary cell}

\ncline{ap0}{ap14}
\ncline{ap1}{ap2}
\ncline{ap3}{ap4}
\ncline{ap5}{ap6}
\ncline{ap7}{ap8}
\ncline{ap9}{ap10}
\ncline{ap11}{ap12}

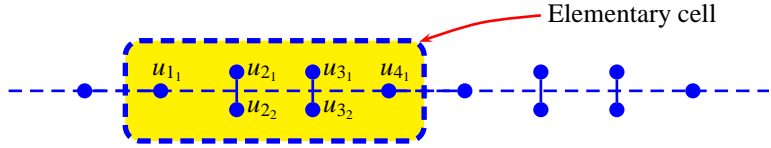
\rput(5,0.2){Fig. 3. An example: the spin chain for the coset ${U(6)\over U(2)\times U(2)\times U(1)\times U(1)}$. }
\end{pspicture}
\end{center}
 
It is convenient to label the unit vectors inside a given unit cell with a double label, where the first-level index shows to which site (inside the elementary cell) the corresponding  vector belongs, and the second-level index shows the position of the vector inside the group at a given site, i.e.
\bea
u_{i_\alpha},\quad\textrm{where}\;\;i=1\cdots m, \;\alpha=1\cdots n_i,
\eea
An example, which clarifies the notation, is shown in Fig. 3.
 
Like we explained above, $\sum\limits_{i=1}^m\,n_i = N$,
so we have $N$ unit vectors $u_{i_\alpha}$ with the following orthogonality properties:
\bea
\bar{u}_{i_\alpha}\circ u_{i_\beta}=\delta_{\alpha\beta}.
\eea
The vectors $u_{j_\alpha}, u_{i_\beta}$ for $i\neq j$ are in generic position. The antiferromagnetic configuration is when \emph{all} of them are orthogonal to each other. The fluctuations around this configuration are conveniently constructed using the Gram-Schmidt (QR) decomposition. We set
\bear
&& z_1=u_{1_1},\cdots\,z_{n_1}=u_{1_{n_1}},\\ \nonumber
&& z_{n_1+1}=u_{2_{1}}+\sum\limits_{s=1}^{n_1}\,a_{\tiny 2_1|1_s}\,u_{1_s},\\ \nonumber && \cdots \\ \nonumber 
&& z_{n_t+q}=u_{t+1_q}+\sum \limits_{m=1}^t\,\sum\limits_{s=1}^{n_m} a_{t+1_q|m_s}\,u_{m_s},\quad 1 \leqslant q\leqslant n_{t+1}
\eear
Summing only over the vectors from previous sites of the spin chain ensures that the $z_{i_\alpha}, z_{i_\beta}$ are orthogonal\footnote{To first order in $a_s$. Generally, one has to write $z_{n_1+2}=u_{2_{2}}+\nu\,u_{2_1}+\sum\limits_{s=1}^{n_1}\,a_{\tiny 2_2|1_s}\,u_{1_s}$ and impose the orthogonality condition $\bar{z}_{n_1+1}\circ z_{n_1+2}=\nu+\sum\limits_{s=1}^{n_1}\,\bar{a}_{\tiny 2_1|1_s}\,a_{\tiny 2_2|1_s}=0$. This shows that $\nu$ is quadratic in the $a$'s and therefore can be neglected in our approximation.}. In what follows we will assume that $a_{i_\alpha|j_\beta}=0$ for $i<j$. The formulas written above are for $z$'s in the same elementary cell. In order to pass to a different cell one has to assign an extra index $k$ in order to be able to differentiate between them. First let us evaluate the interaction between two sites inside one elementary cell:
\bea
i<j \Rightarrow |\bar{z}_{i_\alpha} \circ z_{j_\beta}|^2\sim |a_{j_\beta|i_\alpha}|^2
\eea
The part of the Hamiltonian describing interactions inside block $k$ is as follows:
\bea
H_k= \sum\limits_{i< j}\;d_{j-i}\;\sum\limits_{\alpha, \,\beta}\cdot |a_{j_\beta|i_\alpha}^{\,k}|^2
\eea
Now let us evaluate the interactions between two adjacent blocks, $k-1$ and $k$. We note that
\bea
\Rightarrow |\bar{z}_{i_\alpha}^{(k-1)} \circ z_{j_\beta}^{(k)}|^2\sim |\bar{u}_{i_\alpha}^{(k-1)}\circ u_{j_\beta}^{(k)}+a_{j_\beta|i_\alpha}^{\,k}+\bar{a}_{i_\alpha|j_\beta}^{\,k-1}|^2\sim |\bar{u}_{i_\alpha}^{(k)}\circ \dd_x u_{j_\beta}^{(k)}+a_{j_\beta|i_\alpha}^{\,k}+\bar{a}_{i_\alpha|j_\beta}^{\,k-1}|^2,
\eea
Given our choice of Hamiltonian of interaction range $m-1$, it is easy to show that for $i^{\mathbf{k-1}}<j^{\mathbf{k}}$ the interaction is zero, i.e. $d(i^{\mathbf{k-1}}, j^{\mathbf{k}})=0$ for $i^{\mathbf{k-1}}<j^{\mathbf{k}}$. Indeed, the distance between these sites is then $j^{\mathbf{k}}+(N-i^{\mathbf{k-1}})\geqslant N$, whereas the interaction range is $m-1\leqslant N-1$. Therefore the part of the Hamiltonian corresponding to inter-block interactions can be thus written as
\bea
H_{k-1, k}= \sum\limits_{i>j}\;d_{m-(i-j)}\;\sum\limits_{\alpha, \beta} |\bar{u}_{i_\alpha}^{(k)}\circ \dd_x u_{j_\beta}^{(k)}+\bar{a}^k_{i_\alpha|j_\beta}|^2
\eea
We have used the fact that the function $d(i^{\mathbf{k-1}}, j^{\mathbf{k}})$ depends only on the distance between the sites $i^{\mathbf{k-1}}$ and $j^{\mathbf{k}}$, which, as is easy to see, is the same as $m-(i-j)$ (we are assuming that $i>j$), i.e. $d(i^{\mathbf{k-1}}, j^{\mathbf{k}})=d_{m-(i-j)}$.

The full Hamiltonian has the following form
\bear
&&H=\sum\limits_k\; \sum\limits_{i<j,\,\alpha,\,\beta}\;\left[\,d_{m-(j-i)}\cdot |a_{i_\alpha|j_\beta}^{\,k}|^2 + d_{m-(j-i)}\cdot |a^{\,k}_{i_\alpha|j_\beta}-\bar{u}_{j_\beta}^{(k)}\circ \dd_x u_{i_\alpha}^{(k)}|^2 \right]=\\ \nonumber
&&=\sum\limits_k\; \sum\limits_{i<j,\,\alpha,\,\beta}\;\left[\,(d_{j-i}+d_{m-(j-i)})\cdot |a_{i_\alpha|j_\beta}^{\,k}|^2 -\right.\\ \nonumber
&&\left. - d_{m-(j-i)}\cdot (a^{\,k}_{i_\alpha|j_\beta}\,u_{j_\beta}^{(k)}\circ \dd_x \bar{u}_{i_\alpha}^{(k)}+\bar{a}^{\,k}_{i_\alpha|j_\beta}\,\bar{u}_{j_\beta}^{(k)}\circ \dd_x u_{i_\alpha}^{(k)})+d_{m-(j-i)}\, |\bar{u}_{i_\alpha}^{(k)}\circ \dd_x u_{j_\beta}^{(k)}|^2  \right]
\eear

Now we need to write a corresponding expansion for the kinetic term:
\bea
\mathcal{L}_0= \im\,\sum\limits_{k, \,i<j,\,\alpha,\,\beta}\;\left(\bar{a}_{i_\alpha|j_\beta}^{k} \, \bar{u}_{j_\beta}^{(k)}\circ \dd_t u_{i_\alpha}^{k} -\textrm{c.c.} \right)
\eea
Combining the above expressions we get the full Lagrangian in the form
\bear\label{fulllagr}
&&\mathcal{L}=\sum\limits_k\;\sum\limits_{i<j,\,\alpha,\,\beta}\;\left[\,(d_{j-i}+d_{m-(j-i)})\cdot |a_{i_\alpha|j_\beta}^{\,k}|^2 -\right.\\ \nonumber
&&\left. - 
a^{\,k}_{i_\alpha|j_\beta}\,( d_{m-(j-i)} \, u_{j_\beta}^{(k)}\circ \dd_x \bar{u}_{i_\alpha}^{(k)} +\im u_{j_\beta}^{(k)}\circ \dd_t \bar{u}_{i_\alpha}^{(k)})-\right. \\ \nonumber 
&&\left.-
\bar{a}^{\,k}_{i_\alpha|j_\beta}\,(d_{m-(j-i)}\, \bar{u}_{j_\beta}^{(k)}\circ \dd_x u_{i_\alpha}^{(k)}-\im \bar{u}_{j_\beta}^{(k)}\circ \dd_t u_{i_\alpha}^{(k)})+\right. \\ 
\nonumber &&\left.+ d_{m-(j-i)}\, |\bar{u}_{i_\alpha}^{(k)}\circ \dd_x u_{j_\beta}^{(k)}|^2  \right]
\eear
The last remaining step is to perform Gaussian integration over $a_{i_\alpha|j_\beta}^{\,k}$. This is done using the simple formula
\bea\label{quadint}
A|w|^2+B w +C \bar{w}=A(w+{C\over A})(\bar{w}+{B\over A})-{BC\over A}
\eea
One obtains
\bear
&&\mathcal{L}\to\sum\limits_k\; \sum\limits_{i<j,\,\alpha,\,\beta}\;\left[ \,d_{m-(j-i)}\, |\bar{u}_{i_\alpha}^{(k)}\circ \dd_x u_{j_\beta}^{(k)}|^2-\right. \\ \nonumber &&\left.
-
\frac{1}{d_{j-i}+d_{m-(j-i)})} ( d_{m-(j-i)} \, u_{j_\beta}^{(k)}\circ \dd_x \bar{u}_{i_\alpha}^{(k)} +\im u_{j_\beta}^{(k)}\circ \dd_t \bar{u}_{i_\alpha}^{(k)}) \;\times\right. \\ \nonumber &&\left. \times 
(d_{m-(j-i)}\, \bar{u}_{j_\beta}^{(k)}\circ \dd_x u_{i_\alpha}^{(k)}-\im \bar{u}_{j_\beta}^{(k)}\circ \dd_t u_{i_\alpha}^{(k)}))\right]
\eear
We rewrite it in the form isolating the `metric' part and the $\theta$-term part:
\bear\label{res1}
&&\mathcal{L}\to\sum\limits_k\;  \sum\limits_{i<j,\,\alpha,\,\beta}\;\left[\,\frac{-1}{d_{j-i}+d_{m-(j-i)}}\,\left( |\bar{u}_{j_\beta}^{(k)}\circ \dd_t u_{i_\alpha}^{(k)})|^2-d_{j-i} d_{m-(j-i)} |\bar{u}_{i_\alpha}^{(k)}\circ \dd_x u_{j_\beta}^{(k)}|^2 \right)+\right. \\ \nonumber &&\left.
+ 
\frac{d_{m-(j-i)}}{d_{j-i}+d_{m-(j-i)})} \left( \im\,(  \bar{u}_{j_\beta}^{(k)}\circ \dd_t u_{i_\alpha}^{(k)}) \,(u_{j_\beta}^{(k)}\circ \dd_x \bar{u}_{i_\alpha}^{(k)}) -  \im (u_{j_\beta}^{(k)}\circ \dd_t \bar{u}_{i_\alpha}^{(k)}) (\bar{u}_{j_\beta}^{(k)}\circ \dd_x u_{i_\alpha}^{(k)})\right)
\right]
\eear
As the first line shows, in order for the result to be Lorenz-invariant, we should require
\bea\label{lorenz}
d_{k} d_{m-k}=\textrm{const.}\equiv D^2\quad\textrm{(independent of $k$)}
\eea
In this case the metric on the flag manifold is described in terms of the matrix\footnote{We have rescaled the space coordinate $x\to D x$ in order to set the speed of light equal to 1.}
\bea
\lambda_k=\frac{D}{d_k+d_{m-k}}\, .
\eea 
Another requirement that we impose on our system is that the $\theta$-term is a topological invariant, in other words the second line of (\ref{res1}) should be a closed 2-form. This leads to additional constraints on the coefficients $d_l$. We will get the following result:
\bear
d_k=\sqrt{\frac{m-k}{k}},\;\;\lambda_k=\frac{\sqrt{k\,(m-k)}}{m}\, .
\eear
Indeed, the second line of (\ref{res1}) can be thought of as a pull-back to the worldsheet of the following form:
\bea\label{muk}
\omega=\sum\limits_{i<j}\,\mu_{j-i}\,\omega_{ij}\quad \textrm{with}\;\;\mu_k=\frac{d_{m-k}}{d_k+d_{m-k}}\, ,
\eea
where
\bea
\omega_{ij}=\sum\limits_{\alpha,\beta}\;\im\, \bar{u}_{j_\beta}\circ d u_{i_\alpha} \wedge u_{j_\beta}\circ d\bar{u}_{i_\alpha}\; .
\eea
Introduce also the following \emph{closed} forms
\bea
\omega_i=\sum\limits_{j=1}^m\;\omega_{ij}\, .
\eea
Note that $\omega_i$ represents $c_1(\mathcal{O}_{G_{n_i}}(1))$. There is a single relation between these forms (since $\omega_{ji}=-\omega_{ij}$)
\bea
\sum\limits_{i=1}^m\;\omega_i=0
\eea
We want to find such $\mu_k$'s for which $\omega$ is expressible as a linear combination of $\omega_i$'s:
\bea
\sum\limits_{i=1}^m\,a_i\,\omega_i=\omega=\sum\limits_{i<j}\,\mu_{j-i}\,\omega_{ij}
\eea
One can see that this implies
\bea\label{eqsmu}
\mu_{j-i}=a_j-a_i
\eea
Since the l.h.s. depends only on the difference $j-i$, we find that
\bea
a_j=j \;\alpha+\textrm{const.},
\eea
where $\alpha=\mu_1$ and the constant is inessential, since its effect is to shift the form $\omega$ by $\sum\,\omega_i=0$. Therefore we set the constant to zero. From the explicit expression for $\mu_k$ it follows that $\mu_k+\mu_{m-k}=1$, therefore
\bea
\mu_k+\mu_{m-k}=\alpha\, k+\alpha \, (m-k)=\alpha \, m =1 \Rightarrow \alpha={1\over m}
\eea
In other words
\bea
\omega={1\over m} \sum\limits_{j=1}^m\,j\,\omega_j\,,
\eea
so we have arrived at the result announced in Section \ref{topsec}.

Using the definition of $\mu_k$, (\ref{muk}), and the relations \ref{eqsmu} we get the following equations
\bea
\mu_k=\frac{d_{m-k}}{d_k+d_{m-k}}=\frac{D^2}{d_k^2+D^2}={k\over m}
\eea
Solving for $d_k$, we obtain
\bea
d_k=D\,\sqrt{\frac{m-k}{k}}
\eea
which is the formula reported in (\ref{dk}), up to an inessential factor of  $D$, which is simply the normalization of the Hamiltonian. Clearly, these $d_k$ satisfy the Lorenz invariance condition (\ref{lorenz}). Our derivation is thus complete.

\ssection{Integrating over the flag manifold $\fl_3$}\label{intflag}

In the paper \cite{Bykov} we dealt rather closely with the case when the manifold of coherent states is $\CP^{N-1}$. The reader might wonder, what changes arise in the general case --- the one of the flag manifold. To resolve the doubts we provide an example: namely, we prove the completeness of the system of coherent states
\bea\label{cohstateapp}
\phi_{uvw}(a, b)=(\bar{v}\circ a)\cdot[(\bar{v}\circ a)(\bar{w}\circ b)-(\bar{w}\circ a)(\bar{v}\circ b)],\quad \bar{w}\circ v=0
\eea
in the vector space $V_{\textrm{adj}}$ of the adjoint representation of $\mathfrak{su}_3$. According to (\ref{su3adj}) this space is a subspace of the space of homogeneous polynomials in $(a, b)$ if bidegree $(2, 1)$. On the space of polynomials of degree $m$ in $N$ variables $z_1 \cdots z_N$ we will use the scalar product
\bea\label{scalprod}
\langle f \,|\, g\rangle=\int\,\prod\limits_{i=1}^N\,dz_i\,d\bar{z}_i\,\widebar{f(z)}\,g(z)\,e^{-|z|^2}
\eea
For the case at hand $N=3$, of course. We want to show that for any two states $\langle f|,\;|g\rangle$ the following identity holds:
\bea\label{partunity1}
\langle f \,|\, g\rangle=\int\,d\mu_{\fl_3}(v, w, u)\;\frac{\langle f \,|\, \phi_{uvw}\rangle\, \langle  \phi_{uvw}\,|\, g\rangle}{\langle \phi_{uvw}\,|\,\phi_{uvw}\rangle}
\eea
for some volume element $d\mu_{\fl_3}$ on the flag manifold. Although the coherent state $\phi_{uvw}$ does not depend on the variable $u$ (the third line in the flag), we still need to integrate over it, since it is nontrivially entangled with the other coordinates by the measure. Note that all multiplicative constants arising in the proof may be absorbed in $d\mu_{\fl_3}$, therefore we will not keep track of them. Using (\ref{scalprod}) one finds out that (\ref{partunity1}) is equivalent to the following identity involving the coherent states only:
\bea\label{partunity2}
\int\,d\mu_{\fl_3}(v, w, u)\;\frac{\phi_{uvw}(a, b)\,  \widebar{\phi_{uvw}}(\bar{c}, \bar{d})}{\langle \phi_{uvw}\,|\,\phi_{uvw}\rangle}=\underbrace{(\bar{c}\circ a)\,\left((\bar{c}\circ a)\,(\bar{d}\circ b)-(\bar{c}\circ b)\,(\bar{d}\circ a)\right)}_{\equiv \,\mathbf{Pr}(a, b\,|\,c, d)}
\eea
The reason for this is that the r.h.s. is the `kernel' of the projection operator on the adjoint representation:
\bea
\int \,dc\,d\bar{c}\,dd\,d\bar{d}\;\mathbf{Pr}(a, b\,|\,c, d)\,e^{-|c|^2-|d|^2}\;f(c, d)=f(a, b)\quad\textrm{for}\quad f\in V_{\textrm{adj}}
\eea
The volume element on $\fl_3$ may be written as follows (up to a constant):
\bea\label{volflag}
d\mu_{\fl_3}(v, w, u)= d\mu_{\CP^2}(u)\,d\mu_{\CP^2}(v)\,d\mu_{\CP^2}(w)\,\delta^{(2)}(\bar{w}\circ v)\, \delta^{(2)}(\bar{w}\circ u)\,\delta^{(2)}(\bar{v}\circ u)
\eea
The most convenient way to deal with the $\CP^2$ volume element is to pull it back to $\CC^3$, using the tautological bundle. It can be done as follows:
\bea
d\mu_{\CP^2}(u)=du_1\wedge du_2 \wedge du_3 \wedge d\bar{u}_1 \wedge d\bar{u}_2 \wedge d\bar{u}_3\;e^{-\sum\limits_{i=1}^3\,|u_i|^2}
\eea
We are to integrate functions on $\CP^2$, i.e. functions on $\CC^3$ invariant under a global rescaling. Such functions do not depend on the `radial coordinate' $\sum\limits_{i=1}^3\,|u_i|^2$, therefore the sole reason why we have inserted the Gaussian exponent is to make the integral along this radial direction --- the fiber of the tautological bundle --- convergent. We will also exponentiate the delta-functions in (\ref{volflag}) by means of the standard representation $\delta^{(2)}(\bar{w}\circ v)\sim\int\,d\lambda\,d\bar{\lambda}\;e^{i\,(\lambda\, \bar{w}\circ v+ \bar{\lambda}\, \bar{v}\circ w)}$. Thus, the l.h.s. of (\ref{partunity2}) takes the following form:
\footnotesize
\bear
& \mathcal{I}\equiv \int\,du\,d\bar{u}\,dv\,d\bar{v}\,dw\,d\bar{w}\,\prod\limits_{i=1}^3\, d\lambda_i\,d\bar{\lambda}_i\,\frac{\phi_{uvw}(a, b)\,  \widebar{\phi_{uvw}}(\bar{c}, \bar{d})}{|v|^4\,|w|^2}\times \\ \nonumber  &\!\!\!\!\times \exp{\left[-(|u|^2+|v|^2+|w|^2)+i\,(\lambda_1\, \bar{w}\circ v+ \bar{\lambda}_1\, \bar{v}\circ w+\lambda_2\, \bar{w}\circ u+ \bar{\lambda}_2\, \bar{u}\circ w+\lambda_3\, \bar{v}\circ u+ \bar{\lambda}_3\, \bar{u}\circ v)\right]}\, ,
\eear
\normalsize
where $du\equiv du_1\,du_2\,du_3$ etc. Since the only dependence on $u$ comes from the exponent, it is convenient to integrate over $u$ in the first place. Application of the formula (\ref{quadint}) results in the following expression:
\bear
&\mathcal{I}\sim \int\,dv\,d\bar{v}\,dw\,d\bar{w}\,\prod\limits_{i=1}^3\, d\lambda_i\,d\bar{\lambda}_i\,\frac{\phi_{uvw}(a, b)\,  \widebar{\phi_{uvw}}(\bar{c}, \bar{d})}{|v|^4\,|w|^2}\times\\ \nonumber &\!\!\!\!\!\!\times \exp{\left[-(|v|^2+|w|^2)-|\lambda_2 \bar{w}+\lambda_3 \bar{v}|^2+i\,(\lambda_1\, \bar{w}\circ v+ \bar{\lambda}_1\, \bar{v}\circ w)\right]}
\eear
We notice that $|\lambda_2 \bar{w}+\lambda_3 \bar{v}|^2=|\lambda_2|^2\,|w|^2+|\lambda_3|^2\,|v|^2$, since $\bar{w}\circ v=0$. 
Upon introduction of the variable $t_2=|\lambda_2|^2$ the $w$-integral assumes the form (for the moment we forget about the $v$-integral)\par
\footnotesize
\bea
\int\limits_0^\infty\,dt_2\;\int\,dw\,d\bar{w}\;\; \frac{\phi_{uvw}(a, b)\,  \widebar{\phi_{uvw}}(\bar{c}, \bar{d})}{|v|^4\,|w|^2}\,\exp{\left[-(1+t_2)|w|^2+i\,(\lambda_1\, \bar{w}\circ v+ \bar{\lambda}_1\, \bar{v}\circ w)\right]}
\eea
\normalsize
First of all we integrate by parts with respect to $t_2$ in order to get rid of $|w|^2$ in the denominator. To get rid of the terms in the exponent, linear in $w$, we make a shift $w\to w+\frac{i \,\lambda_1\,v}{1+t_2},\;\bar{w}\to \frac{i \,\bar{\lambda}_1\,\bar{v}}{1+t_2}$. The interesting property of this shift is that it leaves the coherent state (\ref{cohstateapp}) unchanged, due to the fact that $w$ enters (\ref{cohstateapp}) only in an antisymmetric combination with $v$ (see Remark 2 in Section \ref{cohstates}). However, the shift produces an extra term $-{|\lambda_1|^2\,|v|^2\over 1+t_2}$ in the exponent. We introduce the variable $t_1={|\lambda_1|^2\over 1+t_2}$ and integrate by parts twice with respect to $t_1$ to obtain
\bea\footnotesize
\!\mathcal{I}\!\sim\!\!\int\limits_0^\infty\,dt_1\,t_1^2\!\!\int\limits_0^\infty\,dt_2\,t_2\cdot(1+t_2)\,\!\!\int\!\!dv\,d\bar{v}\,dw\,d\bar{w}\; \phi_{uvw}(a, b)\,  \widebar{\phi_{uvw}}(\bar{c}, \bar{d})\cdot
\exp{\left[-(1+t_1)|v|^2-(1+t_2)|w|^2 \right]}
\eea
The inner integral over $v$ and $w$ is Gaussian, to which Wick's theorem is applicable. It can be easily seen to give
\bea
\mathcal{I}\sim\textrm{const.}\;\cdot\; (\bar{c}\circ a)\,\left((\bar{c}\circ a)\,(\bar{d}\circ b)-(\bar{c}\circ b)\,(\bar{d}\circ a)\right)
\eea
Hence we have proven (\ref{partunity2}) up to a constant, that can be absorbed into $d\mu_{\fl_3}$.

\ssection{The quadratic Casimir via oscillator algebra}\label{apposc}

As an example we calculate the value of the quadratic Casimir of $\mathfrak{su}_N$ in the representation described schematically by the following diagram:

\hspace{0.3cm}$\underbrace{\overbrace{\yng(4,3)}^m\!\!\!\!\!\!}_n$\hspace{0.5cm} where we assume there are $m$ boxes in the first row and $n$ boxes in the second one ($m\geqslant n$). We assign $N$ pairs of creation/annihilation operators $\mathbf{a}, \mathbf{a}^\dagger, \mathbf{b}, \mathbf{b}^\dagger$ to each row. The rotation generators are
\bea
S^\alpha=\mathbf{a}^\dagger \circ \tau^\alpha\circ \mathbf{a}+\mathbf{b}^\dagger \circ \tau^\alpha\circ \mathbf{b}
\eea
The generators $\tau^\alpha$ are unit-normalized: $\tr(\tau^\alpha\tau^\beta)=\delta^{\alpha\beta}$. Then $\sum\limits_\alpha\;\tau^\alpha \otimes \tau^\alpha = P-{1\over N} \,I$, where $P$ is the permutation and $I$ the identity operator. Thus, for the Casimir one obtains (here for brevity we omit the state $|\psi\rangle$ on which these operators act, but its presence is implied)
\bear
&&C_2\equiv \sum\limits_\alpha\;S^\alpha\,S^\alpha=\\ \nonumber &&= \overbrace{a_i^\dagger \, a_j\;a_j^\dagger \, a_i}^{=m^2+(N-1)m}+\overbrace{a_i^\dagger \, a_j \;b_j^\dagger \, b_i}^{=-n}+\overbrace{b_i^\dagger \, b_j \, a_j^\dagger \, a_i}^{=-n}+\overbrace{b_i^\dagger \, b_j \; b_j^\dagger \, b_i}^{=n^2+(N-1)n}-
{1\over N}\overbrace{ (a_i^\dagger \, a_i+b_i^\dagger \, b_i)^2}^{=(m+n)^2}=\\ \nonumber &&
= m^2+(N-1)m+n^2+(N-1)n-{1\over N}(m+n)^2-2\,\mathrm{min}(m, n)
\eear
In our case one can replace $\mathrm{min}(m, n)=n$.

\vspace{0cm}

\renewcommand\refname{\begin{center} \centering\normalfont\scshape  References\end{center}}
\bibliography{refs}
\bibliographystyle{ieeetr}

\end{document}